\documentclass{aa}
\usepackage[colorlinks=true,linkcolor=blue,citecolor=blue]{hyperref}%
\usepackage{graphicx}
\usepackage{txfonts}
\usepackage{float}
\usepackage{xcolor}
\usepackage{adjustbox}
\newcommand{\nodata}{$\cdots$}

\begin{document}

   \title{Dark matter density profiles of the Milky Way satellite population: reconciling simulations and observations}

   \author{J. Sarrato-Alós
          \inst{1}\fnmsep\inst{2} \&
          J. M. Arroyo-Polonio
          \inst{1}\fnmsep\inst{2} \thanks{Both authors contributed equally to this work.}
         }

   \institute{Instituto de Astrofísica de Canarias (IAC), Calle Via Láctea s/n, E-38205 La Laguna, Tenerife, Spain\\
   \email{jorge.sarrato@iac.es}
         \and
             Universidad de La Laguna, Avda. Astrofísico Fco. Sánchez s/n, E-38206 La Laguna, Tenerife, Spain
             }

   \date{Received x xx, xxxx; accepted x xx, xxxx}

\abstract
{The cusp-core problem remains a central challenge in the $\Lambda$CDM model. Comparisons between the inner logarithmic slopes ($d\log\rho/d\log r$) of galaxies' dark matter haloes inferred from simulations and observations have historically suffered from a systematic
methodological mismatch. Comparing slopes at different radii raised apparent
tensions between simulations and observations due to inconsistent
slope definitions rather than genuine physical discrepancies.}
{We aim to perform a rigorous comparison between dark
matter density profiles of Milky Way dwarf satellite galaxies inferred
from dynamical modelling and those predicted by CDM cosmological hydrodynamical
simulations in order to evaluate their agreement.}
{We use the NIHAO and FIRE-2 cosmological zoom-in simulation suites 
($M_{\star} \sim 10^3$--$10^{11}\,\rm M_\odot$) to compare simulated dark matter density profiles against 
dynamical model inferences of 16 Milky Way satellites from four 
distinct methods, restricting the analysis to 
their reliability limits. We evaluate simulated slopes at 
observationally accessible radii and directly compare full density and slope 
profiles between simulated and observed satellites of matched stellar mass.}
{Both observed and simulated galaxies reveal a clear mass-dependent
core-formation trend, with inner slopes rising from super-cuspy values
($d\log\rho/d\log r \approx -1.5$) at $M_{\star} \sim 10^5\,\rm
M_\odot$ to core-like values ($\approx 0$ to $-0.4$) at $M_{\star}
\sim 10^8\,\rm M_\odot$. Efficient core formation begins at $M_{\star} \gtrsim 10^6\,\rm M_\odot$, with satellite systems presenting increased scatter at this mass, consistent with tidal disruption. Simulated profiles agree with observations to within $\lesssim
1\sigma$ for most systems, outperforming a NFW profile. Willman\,1 remains an unresolved outlier whose anomalous structural properties are not reproduced by either simulation suite. Observational models present scatter for the same system at a level comparable to the apparent simulation-to-observation offset.}
{Evaluating dark matter density slopes at fixed physical radii significantly improves the agreement between hydrodynamical simulations and dwarf satellites' dynamical models. Moreover, a comprehensive comparison of full density profiles at fixed stellar mass reveals no clear signs of tension.}


   \titlerunning{Reconciling simulated and observed MW satellite density profiles}
   \authorrunning{J. Sarrato-Alós \& J. M. Arroyo-Polonio}

   \maketitle

\section{Introduction}

The distribution of dark matter in galaxies remains a high-priority problem in astrophysics. While cosmological simulations within the $\Lambda$CDM framework predict that dark matter haloes follow a characteristic density profile with a cuspy inner slope \citep[$d\log\rho/d\log r=-1$, ][]{NFW}, observational inferences often suggest shallower, cored profiles, particularly in dwarf galaxies \citep[e.g.,][]{Blok_obs_cores, Oh_2011}. Reconciling this apparent discrepancy (commonly known as the "cusp–core problem") is central to understanding both the nature of dark matter and the interplay between baryonic feedback processes and halo structure. 

On one hand, this discrepancy has motivated the exploration of non-CDM scenarios. These include Warm Dark Matter \citep[WDM; e.g.,][]{Bode2001}, Self-Interacting Dark Matter \citep[SIDM; e.g.,][]{Spergel2000}, and Fuzzy Dark Matter \citep[FDM; e.g.,][]{Hu2000, Hui2017}, each of which can alter inner halo structures through distinct physical mechanisms. WDM suppresses small-scale structure via free-streaming, though it generally requires additional baryonic effects to produce flat inner cores rather than slightly lower-amplitude cusps. In contrast, SIDM features self-scattering that robustly creates a central core in haloes; however, at later times, these systems can undergo core collapse, reverting back to a steep, cuspy profile. FDM introduces ultralight bosons where quantum pressure suppresses structures below the de Broglie wavelength, naturally forming a flat core at the centre of haloes.

On the other hand, over the past decade, hydrodynamical simulations have demonstrated that baryonic processes, in particular stellar feedback, can significantly alter the inner slope of cold dark matter (CDM) density profiles \citep{Governato2010, Pontzen2012}. Several systematic studies, such as those of \citet{DiCintio2014}, \citet{Tollet}, \citet{Lazar} and \citet{SarratoAlos2026}, have quantified the dependence of the inner slope on the stellar-to-halo mass ratio. These works typically measure the slope in simulations within a radial range of 0.01–0.02 $R_{\rm vir}$, and they reveal a strong correlation: haloes hosting galaxies with intermediate stellar-to-halo mass ratios ($M_{\star}/M_{\text{halo}} \sim 5\times10^{-3}$) develop the shallowest dark matter profiles, while both lower and higher mass systems remain cuspier.

On the observational side, constraints on inner dark matter slopes rely primarily on dynamical modelling of stellar and gaseous kinematics. Studies such as \citet{Hayashi_slopes_1, Hayashi_slopes_2}, or \citet{cooke2022} for more massive dwarf galaxies, have attempted to bridge simulations and observations by comparing the predicted slope versus stellar-to-halo mass ratio relation from hydrodynamical CDM runs with the slopes inferred from dynamical models of dwarf galaxies. These comparisons reveal contrasting systematic discrepancies at different regions of the dwarf mass scale. At the lower-mass end ($M_{\star}/M_{\mathrm{halo}} \le 10^{-3}$), where hydrodynamical simulations feature cuspy dark matter profiles, non-spherical Jeans dynamical models indicate inner slopes that are systematically shallower than the simulated relations \citep{Hayashi_slopes_1, Hayashi_slopes_2}. This lower-mass preference for cored structures is matched by gas rotation curve data across the same mass scales \citep{cooke2022}. The latter study also presents hints of the inverse discrepancy at larger stellar-to-halo mass ratios ($M_{\star}/M_{\mathrm{halo}} \sim 10^{-2}$), where some rotation curve fits indicate cuspier slopes than simulations, for which this scale is close to their peak of core formation efficiency.

While this represents an important step towards a direct comparison, we noticed a methodological mismatch: simulations typically report slopes at radii defined as a fixed fraction of the virial radius \citep[$0.01$--$0.02\,R_{\rm vir}$; e.g., ][]{DiCintio2014}, historically labelled as inner slopes, whereas observational analyses can only probe down to the innermost radius accessible with the available data. This mismatch complicates the interpretation of whether apparent discrepancies arise from genuine physical differences or from inconsistent definitions of the inner slope.

In this work, we address this mismatch by adopting a simulation measurement strategy consistent with observational methodology. Rather than evaluating the inner slope at a fixed fraction of the virial radius, we compute it at the specific physical radii accessible to dynamical modelling. This approach ensures a rigorous, apples-to-apples comparison between simulated and observed galaxies. Using this method, we find that this observationally motivated definition significantly improves the agreement between hydrodynamical simulations in a CDM framework and dwarf galaxy observations. Moreover, a comprehensive comparison of full density profiles at fixed stellar mass reveals no clear signs of tension.

Our analysis focuses on the faintest systems with literature models based on fits to resolved stellar kinematics ($M_{\star}\sim10^{3}-10^{8}\,\rm M_\odot$), comparing the observationally inferred dark matter profiles of Milky Way dwarf satellites with predictions from the NIHAO \citep{wang2015} and FIRE-2 \citep{Hopkins_2018} hydrodynamical simulation suites. The proximity of these satellite galaxies allows for resolved stellar kinematics, enabling robust dark matter constraints via various dynamical modelling techniques. However, this low-mass regime introduces distinct challenges: observational data are limited by spatial coverage and potential extrapolation bias, while simulations are bounded by numerical resolution \citep[e.g.,][]{Power2003}. We explicitly account for these systematic limitations to ensure a fair and consistent comparison.

This paper is structured as follows. Sec.~\ref{Sec:literatureestimations} reviews the limitations of existing methods for estimating dark matter density profiles from observations, presenting the observational works with which we compare in this study and defining their reliability radial ranges. Sec.~\ref{sec:Simulations} describes the simulation suites and our post-processing methods to ensure consistency with observations. In Sec.~\ref{sec:results}, we present and discuss our main results, including both the slope-only and direct profile comparisons. Finally, Sec.~\ref{sec:conclusions} summarises our conclusions.

\section{Defining reliable regions from observational inferences}
\label{Sec:literatureestimations}
The inference of dark matter density profiles from observations primarily relies on projected phase-space information. Stellar positions on the sky are typically obtained from photometric surveys \citep[e.g.,][]{GaiaDR3}, while line-of-sight velocities are measured through ground-based spectroscopic observations \citep[e.g.,][]{Tolstoy2006, Walker2009data, Tolstoy2025}\footnote{In a few favourable cases, the combination of \textit{HST} and \textit{Gaia} astrometry enables proper-motion measurements for several hundred stars in some of the most massive Milky Way satellites \citep{Vitral2024}. These data can provide useful additional constraints for dynamical modelling. However, such measurements remain rare, are spatially limited by the field of view of \textit{HST}, and typically contribute less constraining power than line-of-sight velocity samples.}. Samples comprising hundreds, or in rare cases thousands, of stellar line-of-sight velocity measurements are commonly used for dynamical modelling.

Spectroscopic datasets are often affected by significant selection effects arising from limited observing time, fibre-allocation constraints, and target prioritisation strategies. In addition, reliable membership assignment becomes increasingly difficult in the outermost regions, where contamination from foreground stars is more severe and the surface density of member stars is low. Consequently, the effective radial coverage of dynamical tracers can vary substantially across systems. It is essential to assess carefully the radial range over which observational constraints drive the inference of dark matter density profiles, and to determine the spatial scales within which these profiles can be considered robust.

Furthermore, dynamical modelling techniques may introduce additional biases. In particular, methods that rely solely on solving the second-order Jeans equations are affected by the well-known mass–anisotropy degeneracy \citep{Binney1982, Merrit1987}. However, several approaches have been proposed to mitigate this limitation. For example, the presence of multiple stellar populations in some dwarf galaxies can provide additional constraints on the mass profile \citep[e.g.,][]{battaglia2008, zhu2016, Strigari}, while higher-order moments of the velocity distribution, which contain information beyond the velocity dispersion, can also help break the degeneracy \citep{breddels2013, wardana2025, banareshernandez2026}. Furthermore, alternative dynamical modelling techniques, such as distribution-function modelling, fit the full line-of-sight velocity distribution and account for multiple stellar populations, therefore further alleviating this degeneracy \citep{chema-dynmod, Pascale_2026}.

In this work, we compile inferred dark matter density profiles from four primary literature sources: \citet{Hayashi_slopes_1, Hayashi_slopes_2} (H20/H23), \citet{Read_2019} (R19), \citet{chema-dynmod} (AP25), and \citet{Pascale_2025, Pascale_2026} (P25/P26)\footnote{P26 uses different models and datasets for the same galaxies. Throughout this paper, we use their results reported for axisymmetric models fitting two stellar populations with the \citet{W23} dataset.}. We consider only observed galaxies for which at least one simulated galaxy has a comparable stellar mass, defined as being within $\pm 0.25$ dex in stellar mass, for reference, the lowest-mass galaxy in our combined FIRE-2 and NIHAO sample has $\log_{10} M_\star/M_{\odot} = 3.16$. Significant modelling differences exist among these distinct analysis frameworks. However, robustly accounting for which specific technique is most reliable within each region of a galaxy is exceptionally challenging and remains beyond the scope of this paper. We therefore assume that these various techniques all provide meaningful constraints within radial regions sufficiently populated by spectroscopic tracers. 

Under this assumption, the primary limitation is established by the data quality rather than by the modelling framework itself. Therefore, for each individual system, we must define objective criteria to determine the radial range over which the inferred profile is genuinely constrained by observations, as opposed to being driven by extrapolation or modelling priors. To accomplish this, we analyse directly the spectroscopic catalogues used in each literature work. We define the reliable radial range using distinct criteria depending on whether the dynamical analysis fits individual stars or binned velocity dispersion data\footnote{The adopted thresholds should be regarded as practical working definitions rather than rigorous theoretical limits. There is no unique minimum number of stars or velocity-dispersion measurements required to constrain a dark matter density slope, as the available dynamical information depends on the radial distribution of the stars, the velocity uncertainties, and the applied technique.}.

For star-by-star analyses, we always consider the profile at the half-light radius as reliable. The inner reliability limit is determined by the smallest radius enclosing at least 50 member stars with line-of-sight velocity measurements. The projected half-light radius is computed as $R_h = R_{h,e}\sqrt{1-e}$, where $R_{\rm h,e}$ is the projected half-light radius along the semi-major axis and $e$ is the ellipticity of the galaxy. If the radius enclosing 50 stars exceeds $R_{\rm h}$, the half-light radius is adopted as the inner limit instead. The outer reliability limit is set by the radius beyond which fewer than 10 member stars are available.

For analyses based on binned velocity dispersion data, we require a minimum of three bins to constrain a slope. The radius of the third innermost bin therefore defines the inner reliability limit. We consider these models reliable all the way out to the outermost bin. These binned-data criteria are necessarily less restrictive than those applied to discrete analyses. This choice is consistent with the fact that individual stellar measurements retain more information than binned summaries. We note that the reliability limits are derived from projected quantities and subsequently applied to three-dimensional profiles. This approximation is unavoidable given that the line-of-sight position of individual stars is unknown.

\section{Simulations}
\label{sec:Simulations}

\subsection{NIHAO}
\label{sec:NIHAO}

The NIHAO project \citep[Numerical Investigations of Hundred Astrophysical Objects][]{wang2015} comprises a suite of high-resolution zoom-in hydrodynamical simulations. These simulations utilise the {\sc gasoline2} code \citep{wadsley2017}, and a Planck cosmology \citep{planckcosmo} framework: H$_{\text{0}}$ = 100h km s$^{\text{-1}}$ Mpc$^{\text{-1}}$ with h = 0.671, $\Omega_{\text{m}}$ = 0.3175, $\Omega_{\Lambda}$ = 0.6824, $\Omega_{\text{b}}$ = 0.049, and $\sigma_{\text{8}}$ = 0.8344.

Galaxy formation is modelled via the MaGICC framework \citep[Making Galaxies In a Cosmological Context;][]{stinson13}, which has demonstrated success in reproducing observed scaling relations, including the Tully-Fisher, size-luminosity, and mass-metallicity relations, \citep{brook12b,maccio2012}. Star formation follows a \citet{chabrier03} initial mass function, occurring in gas regions exceeding a number density threshold of $n_{\rm th} = 10.3\,\rm cm^{-3}$. Stellar feedback incorporates both early energy injection from massive stars and supernova blast-waves \citep{stinson06}. Gas experiences metal-line cooling, photoionisation, and ultraviolet heating, based on the prescriptions detailed by \citet{shen}.

The simulation resolution (100–800 pc) allows for resolving internal structures below 1\% of the virial radius. We used AHF \citep[Amiga Halo Finder;][]{ahf} to identify haloes within each zoom-in volume. From the AHF catalogue, we selected haloes containing more than 99\% high-resolution particles ({\tt fMhires} $>$ 0.99) and at least 50 star particles, both globally and within the inner 0.1\,$R_{\rm vir}$. This limit is set empirically to ensure baryonic properties like stellar masses and half-light radii are well resolved, while accounting for possible contamination from nearby galaxies. Haloes are classified as centrals or satellites based on the AHF {\tt hostHalo} flag. Our final NIHAO sample comprises 425 haloes drawn from 95 zoom-in simulations, including 326 centrals and 99 satellites.

\subsection{FIRE-2}
\label{sec:FIRE}

The Feedback In Realistic Environments \citep[FIRE][]{Hopkins2014} project consists of a series of cosmological zoom-in simulations produced using the meshless finite-mass code GIZMO \citep{gizmo}. The suite uses slightly varying cosmological parameters across different runs. Some align with Planck results \citep{planckcosmo}: H$_{\text{0}}$ = 100h km s$^{\text{-1}}$ Mpc$^{\text{-1}}$ with h = 0.671, $\Omega_{\text{m}}$ = 0.3175, $\Omega_{\Lambda}$ = 0.6824, $\Omega_{\text{b}}$ = 0.049, and $\sigma_{\text{8}}$ = 0.8344. Other simulations adopt the AGORA project parameters \citep{AGORA_cosmo}: H$_{\text{0}}$ = 100h km s$^{\text{-1}}$ Mpc$^{\text{-1}}$ with h = 0.702, $\Omega_{\text{m}}$ = 0.272, $\Omega_{\Lambda}$ = 0.728, $\Omega_{\text{b}}$ = 0.0455, and $\sigma_{\text{8}}$ = 0.807.

Specifically, we use the FIRE-2 simulations \citep{Hopkins_2018}, which include a detailed model of galaxy formation physics. The core zoom-in runs implement stellar feedback channels that include the energy, momentum, mass, and metal fluxes from supernovae (Types Ia and II), continuous stellar mass-loss via O-star and AGB winds, radiation pressure, photoionisation, and photoelectric heating. Furthermore, star formation follows a \citet{Kroupa2002} initial mass function and is strictly restricted to cold, self-gravitating, and molecular gas that exceeds a high critical density threshold ($n_{\rm H} \ge 1000\,\rm cm^{-3}$), where the molecular gas fraction is determined dynamically by accounting for self-shielding. These simulations reach high spatial resolution, with gravitational softening lengths on the order of 10 pc.

The FIRE-2 suite has been extensively validated against observations, successfully reproducing realistic star formation histories, gas distributions, metallicity profiles, morphologies, and the macroscopic features of dwarf and Milky Way-mass rotation curves, as well as stellar mass scaling relations \citep{Hopkins_2018}. While these simulations successfully capture broad kinematic properties on global scales, our work extends this analysis by conducting a rigorous comparison of the smaller-scale inner density slopes against individual local dwarf satellite systems.

We used the Rockstar halo catalogues \citep{Behroozi_2012}, which are publicly available for each zoom-in volume. From the Rockstar catalogues, we retained haloes containing a minimum of 50 star particles within 0.1\,$R_{\rm vir}$. These catalogues do not directly flag haloes as either centrals or satellites; we distinguish them by iterating through the catalogue in order of decreasing mass: a halo is flagged as a satellite if its centre lies within the virial radius of a more massive halo\footnote{This instantaneous geometric cut may introduce minor contamination in the sample of centrals from exited splashback systems \citep{Adhikari_2014}.}. As an additional quality cut, we required the offset between the shrinking-sphere centre \citep{Power2003} and the Rockstar-reported halo centre to be less than 1\,kpc, filtering out haloes with poorly converged centres. Our final FIRE-2 sample consists of 573 haloes drawn from 26 zoom-in simulations, comprising 294 centrals and 279 satellites.

\subsection{Comparison to observations}
\label{sec:sim-vs-obs}
\begin{figure}[htb]
    \centering
    \includegraphics[width=\linewidth]{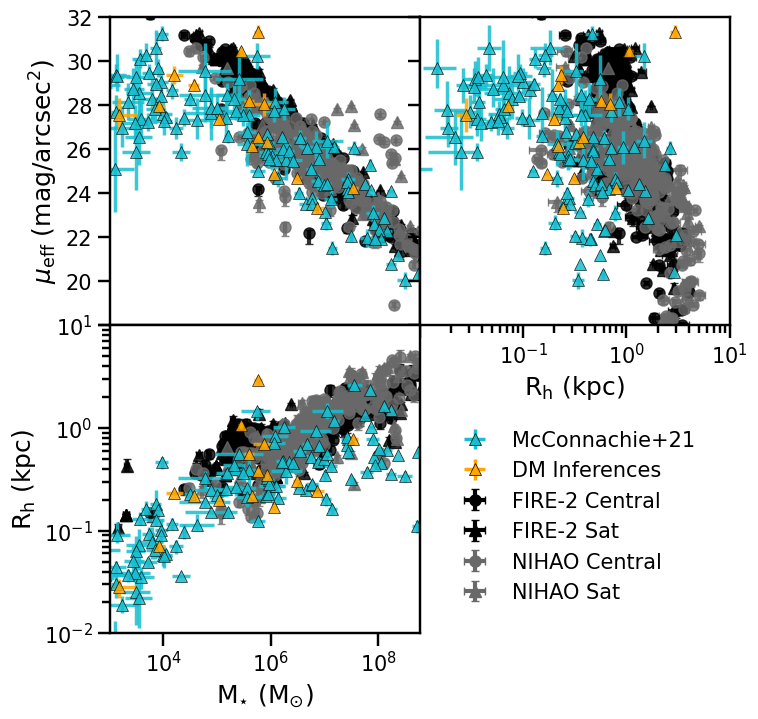}
    \caption{Comparison of the properties of NIHAO (grey) and FIRE-2 (black) isolated (circles) and satellite (triangles) galaxies to Local Group galaxies from the updated \citet{McConnachieNearbyWeb} catalogue (coloured triangles). Top-left panel: Effective surface brightness plotted against stellar mass. Top-right panel: Effective surface brightness against circularised projected half-light radius. Bottom panel: Circularised projected half-light radius against stellar mass. Orange triangles indicate dwarf satellite galaxies analysed in R19, H20/H23, AP25, and P25/P26, while blue triangles represent the rest of the observational samples.}
    \label{fig:sim-obs}
\end{figure}

We compare the baryonic structural parameters (stellar mass, circularised projected half-light radius, and effective surface brightness) of simulated galaxies in our sample with observations of dwarf galaxies, including those with dynamical models from R19, H20/H23, AP25, and P25/P26. Observational data are extracted from the publicly available \citet{McConnachieNearbyWeb} database, which tracks literature updates on the original work presented in \citet{McConnachie2012}. With respect to the galaxies for which we compiled dynamical models, data for Sextans and Canes Venatici\,I are extracted from the original \citet{McConnachie2012}, Crater\,2 from \citet{Cra2Fig1}, Antlia\,II from \citet{AntliaIIFig1}, and all remaining systems from \citet{Munoz2018}. Stellar masses are computed assuming a V-band mass-to-light ratio of 1.68. This value is derived from a synthetic stellar population based on BaSTI evolutionary models \citep[a Bag of Stellar Tracks and Isochrones,][]{Pietrinferni2013} and a Salpeter initial mass function \citep{Salpeter1955}, assuming a constant star formation history between 12 and 13~Gyr ago and a metallicity distribution centred at $-2.3$ with a spread of $0.5$~dex. Fig.~\ref{fig:sim-obs} shows the joint distributions of these parameters. For simulated galaxies, we calculate the projected half-light radius and effective surface brightness in the V-band\footnote{For this calculation, we make use of the publicly available pynbody code \citep{pynbody}. It assigns luminosities to each stellar particle using simple stellar populations from \citet{Marigo2008, Girardi2010}.} from 256 projections uniformly distributed along the unit sphere, and plot the median and $1\sigma$ range.

Simulations exhibit general agreement with observational trends. However, we note that, for stellar masses higher than $\sim$10$^{7}$M$_{\odot}$, the half-light radii of simulated galaxies are systematically larger than those of field galaxies in \citet{McConnachie2012}, closer to those of ultra-diffuse galaxies \citep[e.g.,][]{van_Dokkum_2015}. Other recent studies from the literature find similar trends. \citet{Merritt2020} report light profiles of galaxies in the Illustris TNG100 \citep{Nelson2017} simulations are more extended than observations from DNGS \citep[Dragonfly Nearby Galaxies Survey,][]{Merritt2016}. Furthermore, \citet{Rohr2022} compared FIREbox \citep[][which uses the same galaxy formation model as FIRE-2 simulations in this work]{Feldmann2023} simulated galaxies with galaxy size-stellar mass relations obtained by \citet{Nedkova2021}, and also concluded that the sizes of simulated galaxies are larger than their observational counterparts at the same stellar masses.

\subsection{Inner slopes of simulated galaxies}
\label{sec:sim-inner-slopes}

The inner slope of dark matter density profiles in cosmological simulations is frequently reported at $\sim$1.5\% $R_{\rm vir}$ \citep[e.g.,][]{DiCintio2014, Tollet, Lazar, SarratoAlos2026}. However, this convention has significant limitations. First, measurements at this radius often coincide with the profile's slope transition region, making them unrepresentative of the asymptotic inner slope. Second, $R_{\rm vir}$ is generally inaccessible in observational studies. Finally, for satellite galaxies (the primary targets for dynamical modelling) the virial radius is not well defined due to tidal stripping \citep[e.g.,][]{Read2006, Jiang2013}.

\begin{figure}[htb]
    \centering
    \includegraphics[width=\linewidth]{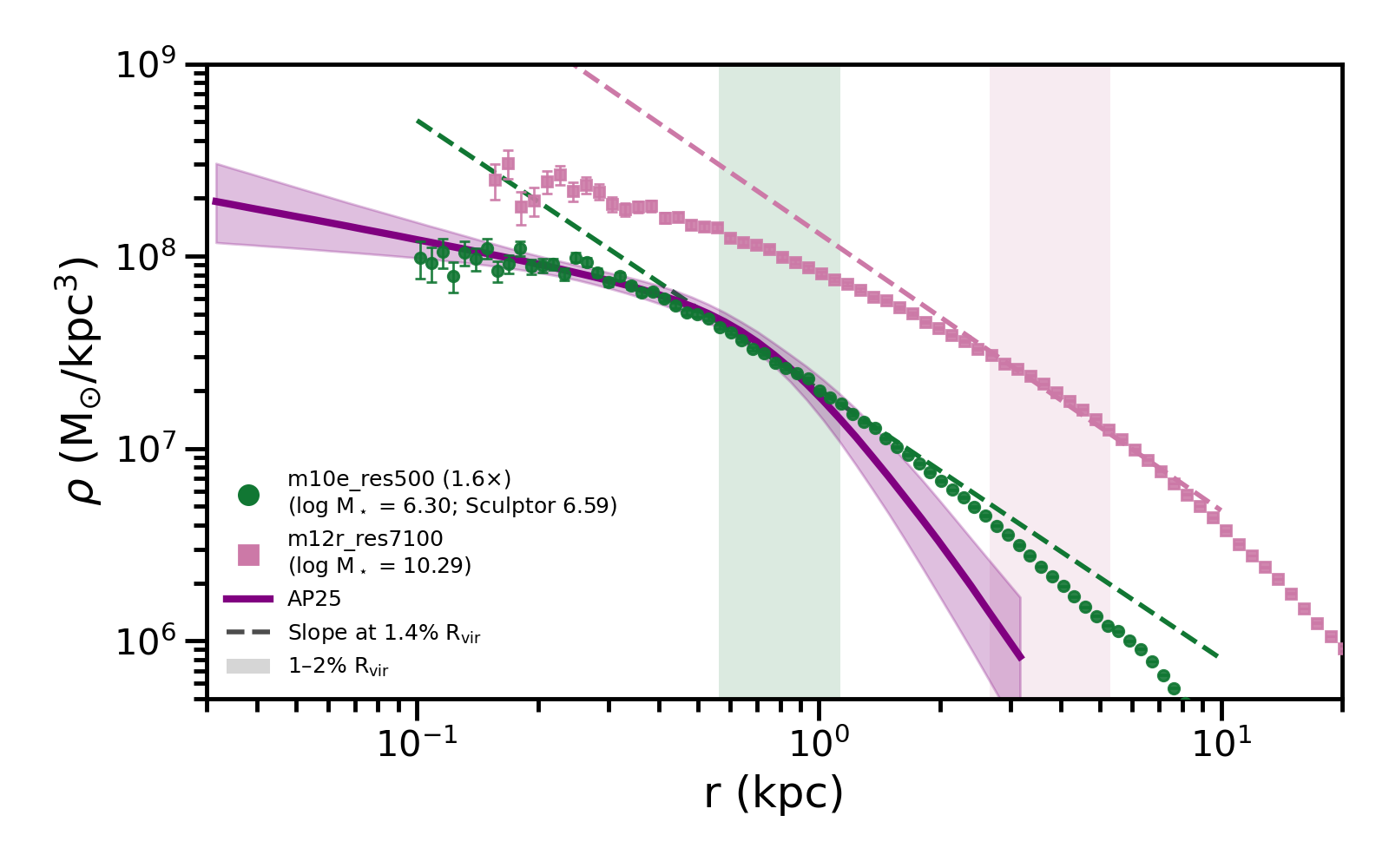}
    \caption{Dark matter density profiles for two galaxies from the FIRE-2 project, shown as points. The shaded regions indicate the 1-2\% $R_{\rm vir}$ range, with the dashed lines representing the slope at the logarithmic midpoint. The simulated galaxy m10e\_res500 is selected to match the stellar mass of the Sculptor dwarf galaxy; its density is scaled by 1.6 to facilitate the visual comparison of profile shapes with the dynamical analysis by AP25. The simulated galaxy m12r\_res7100 is shown as a representative example of a galaxy with a higher stellar mass}
    \label{fig:profiles_slopes_rvir}
\end{figure}

Fig. \ref{fig:profiles_slopes_rvir} shows dark matter density profiles for two FIRE-2 galaxies. While the slope measured at 1–2\% $R_{\rm vir}$ is commonly used as a proxy for the inner slope, the profiles continue to flatten significantly interior to this region. Using this definition to compare with dynamical models of observed dwarf galaxies can be misleading. For example, the figure compares the simulation profile with the AP25 model of Sculptor. The slope inferred at 1–2\% $R_{\rm vir}$ appears cuspier than a more representative inner slope measured at smaller radii, even though the simulation resolves the inner region and such a measurement is feasible. Comparing both the simulated galaxy and the dynamical model at the same physical radii would provide a fairer assessment.

Based on visual inspection, we find that an inner range of 100–500 pc is typically sufficient to avoid the slope transition region. To improve the robustness of comparisons between simulations and observations, we therefore choose to estimate inner slopes at the same physical radius in both.

\subsubsection{Determining the slope profile}
We aim to fit the dark matter density profile of each halo with an analytical function that allows calculating the logarithmic slope at any radius. For this, we adopt the double power-law profile \citep{zhao1996}:

\begin{equation}
\label{eq:dpl}
\rho (r) = \frac{\rho_{s}}{\left( r/r_{s}\right)^{\gamma}\left[ 1 + \left( r/r_{s}\right)^{\alpha}\right]^{\frac{\beta-\gamma}{\alpha}}}\,,
\end{equation}

where $\rho_{s}$ and $r_{s}$ are the characteristic scale density and radius, respectively. We construct dark matter density profiles using 100 logarithmically spaced spherical bins, extending from the radius enclosing the innermost 100 dark matter particles to $R_{\rm vir}$. We define $R_{\rm vir}$ as the radius enclosing an average density $\Delta_{\rm vir}$ times the critical density of the universe, $\rho_{\rm crit} = 3H^{2}/8\pi G$, where $\Delta_{\rm vir}$ follows the redshift-dependent definition by \citet{virial_overdensity} evaluated with the $\Omega_{\rm m}$ value of each simulation.

Before fitting, we apply a radial mask to ensure numerical reliability. In the inner regions, we adopt simulation-specific convergence criteria. For the FIRE-2 simulations, we exclude data below $r_{\rm 200p}$, defined as the radius enclosing the innermost 200 dark matter particles. \citet{Hopkins_2018} validated this threshold by comparing fiducial-resolution runs with higher-resolution resimulations, finding that the enclosed dark matter density profile converges to within $\sim$20\% at this radius. For the NIHAO simulations, where such resolution studies are not available, we instead apply the \citet{Power2003} relaxation-time convergence criterion with the revised threshold of \citet{Schaller2015}. Specifically, we compute the convergence parameter

\begin{equation}
f(r) = \frac{\sqrt{200}}{8}\sqrt{\frac{4\pi\rho_{\rm crit}}{3m_{\rm DM}}}\frac{\sqrt{N(<r)}}{\ln N(<r)}r^{3/2}\,,
\end{equation}

where $N(<r)$ is the number of dark matter particles enclosed within radius $r$ and $m_{\rm DM}$ is the particle mass. We take the convergence radius as the smallest $r$ at which $f(r) \ge 0.33$, corresponding to $\sim$20\% convergence in density \citep{Schaller2015}. Data below this radius are excluded from the fit.

Hereafter, we will refer to these different definitions jointly as $r_{\mathrm{conv},20}$, indicating the radius where density profiles converge to $\sim$20\%.

\subsubsection{NFW Baseline Comparison}
\label{sec:NFWbaseline}

The standard Navarro-Frenk-White \citep[NFW;][]{NFW} profile represents a specific case of the general double power-law configuration presented in Eq.~\ref{eq:dpl}, corresponding to the parameter choice $(\alpha, \beta, \gamma) = (1, 3, 1)$, and is expressed as:

\begin{equation}
\label{eq:nfw_equation_p4}
\rho_{\rm NFW}(r) = \frac{\rho_{s}}{\left(\frac{r}{r_{s}}\right)\left(1 + \frac{r}{r_{s}}\right)^{2}}\,.
\end{equation}

To establish a theoretical baseline, we construct NFW proxy profiles as a function of stellar mass through abundance-matching. We create a fine grid of stellar masses and, for each grid point, we invert the \citet{Moster_2013} stellar-to-halo mass relation to obtain the corresponding $M_{\rm 200c}$, from which we derive $R_{\rm 200c}$ using the critical overdensity definition. The NFW scale radius $r_s$ and concentration $c$ are then assigned via the \citet{Moline_2023} relation for subhaloes, since we aim to compare with dynamical models of Milky Way satellites.

For each stellar mass along the grid, this yields an NFW proxy profile characterised by $(r_s, c, R_{\rm 200c})$. Evaluating the analytic NFW logarithmic slope at any physical radius provides a dark-matter-only NFW expectation for the inner density slope as a function of stellar mass, against which the slopes measured in hydrodynamical simulations or inferred from observations can be directly compared at identical physical scales.

To propagate the uncertainty in abundance matching into the NFW baseline, we perform a Monte Carlo sampling of the \citet{Moster_2013} relation parameters, drawing from their reported $1\sigma$ uncertainties. Each draw produces a different NFW profile, resulting in scatter bands for the inner slope at a fixed stellar mass that reflect the propagated uncertainty from the stellar-to-halo mass relation into the NFW prediction.

\section{Results}
\label{sec:results}

We present the comparison of the dark matter profiles inferred with dynamical models in R19, H20/H23, AP25, and P25/P26 against those from our simulation dataset. We structure this comparison in two complementary ways. First, we evaluate the inner logarithmic slope at three fixed physical radii (100, 250, and 500\,pc) as a function of stellar mass, comparing simulations, observations, and the NFW baseline. Second, we perform a direct profile-by-profile comparison for each observed galaxy, both in density and slope space, which we further extend for the primary satellite-only sample in Appendix~\ref{sec:appendix-profiles} and for the sample of simulated centrals in Appendix~\ref{sec:appendix-centrals}.

\subsection{Inner slope versus stellar mass at fixed physical radius}
\label{sec:slope-vs-mstar}

\begin{figure*}
    \centering
    \includegraphics[width=\textwidth]{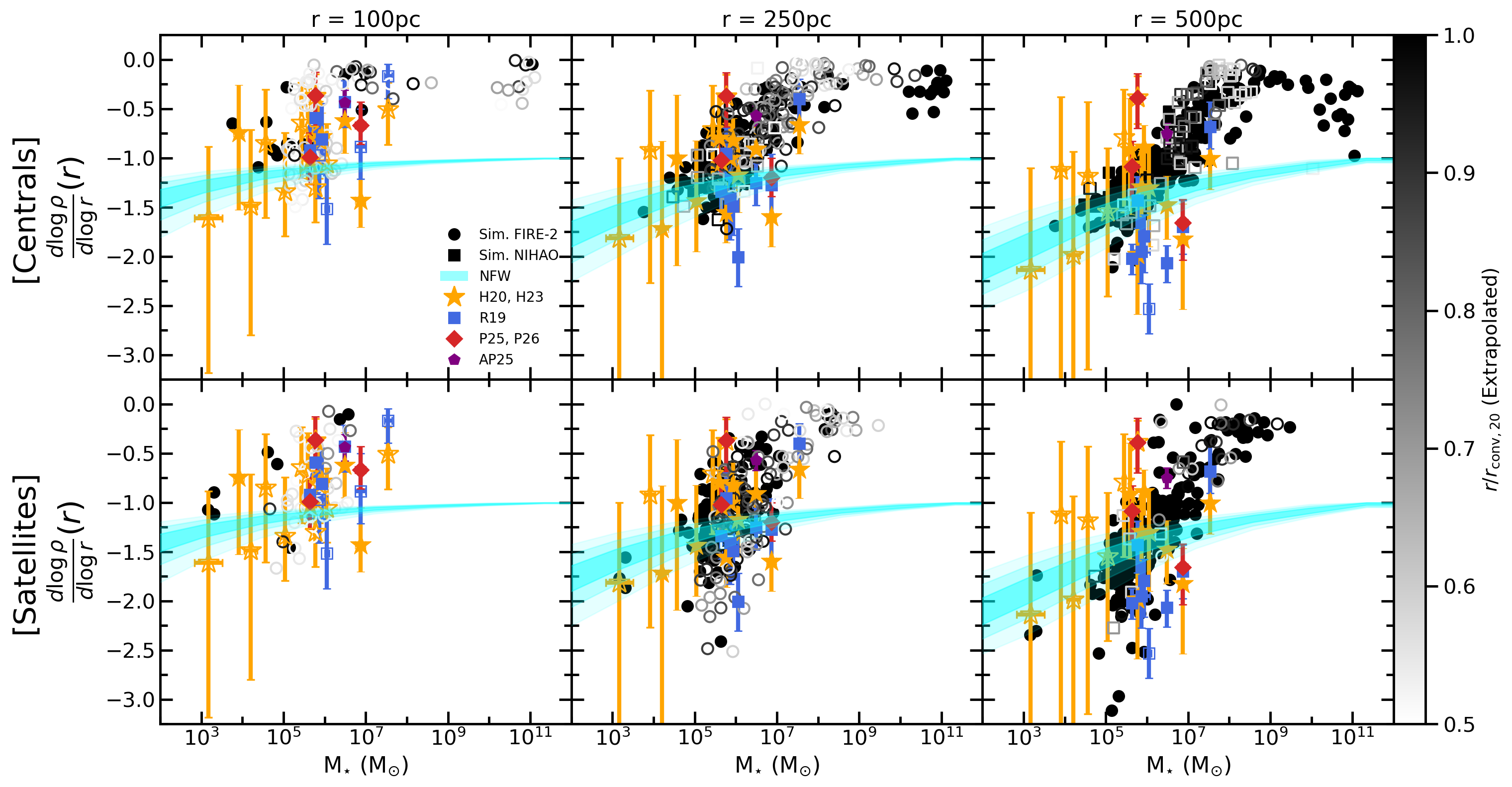}
    \caption{Dark matter density slope $d\log\rho/d\log r$ evaluated at three fixed physical radii versus stellar mass. Each column corresponds to a different radius: 100\,pc (left), 250\,pc (middle), and 500\,pc (right). The upper row shows central (isolated) simulated galaxies, the lower row shows satellites, as indicated in the rows. Grey scale symbols represent simulated galaxies: circles for FIRE-2 and squares for NIHAO. Solid black symbols indicate resolved galaxies, while empty symbols represent unresolved galaxies whose density profile fit we extrapolate up to half of the convergence radius, at maximum. Lighter colour symbols indicate cases of further profile extrapolation. Observational estimates from H20/H23 (orange stars), R19 (blue squares), AP25 (purple pentagon), and P25/P26 (red diamonds) are overlaid with their 68\% credible intervals, and they are also represented as empty markers when the studied radius falls outside their reliability range (see Sec. \ref{Sec:literatureestimations}). The cyan shaded bands show the 1$\sigma$, 2$\sigma$, and 3$\sigma$ scatter of NFW profiles derived from the \citet{Moster_2013} stellar-to-halo mass relation combined with the \citet{Moline_2023} concentration-mass relation, propagated via Monte Carlo sampling of the \citet{Moster_2013} parameter uncertainties.}
    \label{fig:comparison_slopes_hlr}
\end{figure*}

Fig.~\ref{fig:comparison_slopes_hlr} shows the logarithmic dark matter density slope $d\log\rho/d\log r$ evaluated at 100, 250, and 500\,pc as a function of stellar mass, separately for central (isolated) galaxies (upper row) and satellites (lower row). Tables \ref{tab:slope_100pc}, \ref{tab:slope_250pc} and \ref{tab:slope_500pc} (Appendix~\ref{sec:appendix-slope-tables}) indicate the logarithmic slope values derived from the dynamical models of observed satellite galaxies at 100\,pc, 250\,pc and 500\,pc, respectively. It is important to note that, as the galaxies span a wide range of stellar masses and sizes, these fixed radii probe a different region relative to the galaxy’s half-light radius. To identify how inner these distances are, we also assess these radii relative to the virial radius: 100 and 250 pc lie within 1.5\%$R_{\rm vir}$ for most systems considered and can be regarded as measurements of the inner slope in the sense commonly used in the literature. The 500 pc radius meets this criterion only for galaxies with $M_\star>10^6{\rm M}_\odot$. These radii are selected particularly to probe the innermost regions that can be reliably resolved in both simulations and observations across the stellar-mass range considered.

\subsubsection{Core formation and its mass dependence}

For isolated galaxies evaluated at 500\,pc, the simulations reveal a clear trend: inner slopes rise from super-cuspy values of $\approx -1.5$ at $M_{\star} \sim 10^5\,\rm M_\odot$ to shallow, core-like values of $d\log\rho/d\log r \approx 0$ to $-0.4$ at $M_{\star} \sim 10^8\,\rm M_\odot$, before steepening moderately again at the highest masses probed ($M_{\star} \gtrsim 10^{10}\,\rm M_\odot$). This behaviour is consistent with the baryonic feedback-driven core formation mechanism identified in previous work \citep{DiCintio2014, Tollet, Lazar,SarratoAlos2026}, where repeated cycles of star formation and supernova-driven outflows flatten the inner dark matter cusp most efficiently at intermediate stellar-to-halo mass ratios. The steepening at high stellar masses is consistent with the regime where the deep gravitational potential resists feedback-driven fluctuations and adiabatic contraction begins to dominate; although historically it has only been reported when studying the logarithmic slope at fixed fractions of $R_{\rm vir}$, never at fixed physical scales.

At 250\,pc, the same trend is visible, albeit with some differences. At this radius, galaxies generally present more cored slope values, as expected from this more inner region. This also causes the peak of core formation to be an extended range of stellar masses of $\sim 10^8$-$10^{10}\,\rm M_\odot$, rather than a well-defined value over which a sudden re-steepening occurs, as seen traditionally in $R_{\rm vir}$-based studies. At 100\,pc, the apparent shallowing is most extreme, with a large fraction of galaxies at $M_{\star} \gtrsim 10^7\,\rm M_\odot$ approaching slopes $\approx 0$. However, the majority of these points correspond to galaxies for which the evaluation radius lies below the simulation convergence radius $r_{\rm conv,20}$. Although we are conservative with the extent of extrapolation, the 100\,pc column should be interpreted as providing only a qualitative indication of the direction of the trend. We therefore base most of our quantitative statements primarily on the 250 and 500\,pc columns, where the fraction of converged profiles is substantially higher and the slope estimates are more reliable.

A critical mass scale for core formation is evident at both most reliable radii: below $M_{\star} \approx 10^{5.5}\,\rm M_\odot$ many simulated slopes fall within or below the NFW prediction bands, indicating that stellar feedback is insufficient to transform a cusp into a core in this regime. This is broadly consistent with theoretical work \citep[e.g.,][]{Penarrubia2012, Brook2015} showing that feedback-driven core formation becomes most efficient at much higher stellar masses, around $M_{\star}\sim 10^{8.5}\,\rm M_{\odot}$, and that core formation becomes increasingly difficult in fainter dwarfs. We note that the transition is gradual rather than sharp, with non-negligible scatter in both directions; nevertheless, the mass scale $M_{\star} \sim 10^6\,\rm M_\odot$ represents a robust estimate for the lower limit of efficient core formation in these simulations.

\subsubsection{Centrals versus satellites}
Satellite galaxies (lower row of Fig.~\ref{fig:comparison_slopes_hlr}) occupy broadly the same region of the $M_{\star}$-$d\log\rho/d\log r$ plane as isolated galaxies, but with systematically larger scatter. In particular, at 250 and 500\,pc, the satellite distribution develops a pronounced tail towards steep slopes ($\approx -2.5$ to $-3$) with no counterpart in the central galaxy sample. This excess of steep slopes in satellites may be related to gravitational tides, which preferentially disrupt the mass distribution by stripping material from the outer regions of the subhaloes. When we evaluate profiles at identical physical distances, this structural disruption may affect galaxies differently across mass scales. In particular, lower-mass systems typically have shallower potential wells, meaning that a fixed physical radius probes regions further out relative to their scale radii. Consequently, the mass distribution at these fixed scales is far more vulnerable to tidal effects than in more massive counterparts \citep{Penarrubia2010, Green2019}. A detailed investigation of how the dark matter density slopes of individual simulated satellites evolve along their orbits under the influence of tidal effects, and how this evolution compares with the slopes inferred for observed galaxies, would provide a valuable test of this interpretation and represents an interesting avenue for a separate contribution.

At a fixed stellar mass, our sample includes satellites with a diverse range of central halo masses and unique orbital histories. These environmental factors control the degree to which an individual system is affected by tides, increasing the scatter of the dark matter density profile slopes at fixed physical scales. Under this interpretation, the split observed in the slopes of the satellite population at stellar mass scales where isolated galaxies uniformly exhibit profiles flatter than NFW ($M_{\star} \approx 10^6$--$10^7\,\rm M_\odot$), while a fraction of the satellite sample presents NFW-like or even cuspier inner slopes, is entirely consistent with varying degrees of tidal disruption.

Since all observed targets in our comparison are Milky Way satellite galaxies, the satellite row provides the appropriate theoretical reference. The wider scatter in the satellite sample means that a broader range of $d\log\rho/d\log r$ values is physically expected at fixed stellar mass, which has direct implications for the interpretation of apparent simulation to observation discrepancies: a slope difference that would be anomalous for an isolated galaxy may fall within the normal range of the satellite population. In fact, satellite galaxies in our sample provide better matches than centrals for the steepest dynamical models in the literature, such as those of R19 (blue squares) around $M_{\star} \sim 10^6\,\rm M_\odot$.

\subsubsection{Comparison with observations and the NFW baseline}

The observational estimates occupy the same region of the $M_{\star}$-$d\log\rho/d\log r$ plane as the simulated satellites at all three evaluation radii, with no systematic offset that persists across all datasets, although our dataset includes very few galaxies below $M_{\star} = 10^5\,\rm M_\odot$. However, the four sets of dynamical models do not always agree with one another to within their quoted uncertainties for all systems, which complicates any statement about overall simulation-observation consistency.

The H20/H23 measurements (orange) carry the largest individual uncertainties, often spanning $\Delta d\log\rho/d\log r \sim 1$-$1.5$ at 1$\sigma$, and are therefore consistent with both cusps and cores across the entire stellar mass range sampled. This is expected especially because their studies are the only ones including the faintest ($M_{\star} < 10^5\,\rm M_\odot$) dwarf galaxies, for which stellar kinematic data are very limited. However, their uncertainties are still slightly larger than the rest for higher stellar masses. The R19 estimates (blue) are the most cusp-like of the four datasets, clustering near or within the NFW prediction bands, with slopes of $d\log\rho/d\log r \approx -1.5$ to $-2$ at 500\,pc, except the more cored point at higher stellar mass, corresponding to the Fornax dwarf spheroidal galaxy. The AP25 estimate (purple) consistently indicates a somewhat cored distribution for the Sculptor galaxy at all studied physical scales. Finally, the P25/P26 results (red) cluster around $d\log\rho/d\log r \approx -0.4$ to $-0.7$ at 100-250\,pc, finding both a well-developed core, a NFW consistent profile, and a mixed result for their three studied satellites.

The NFW prediction bands (cyan shading) provide a useful dark-matter-only reference. At 500\,pc, the NFW slope ranges from $d\log\rho/d\log r_{\rm NFW} \approx -1.75$ at $M_{\star} \sim 10^5\,\rm M_\odot$ to $\approx -1.25$ at $10^7\,\rm M_\odot$. This trend reflects the weak mass dependence from the \citet{Moline_2023} concentration relation. Simulated galaxies in the peak core-forming mass range lie at $\Delta d\log\rho/d\log r \approx 0.7-1.0$ shallower than this baseline. This departure is statistically significant given the narrow width of the NFW bands. Cored observational points sit well above the NFW expectation. This group includes the AP25 profile and the most cored P25/P26 marker. Generally, we find no significant tension between simulations and observations. They span the same regions of the plane, except for the 500\,pc evaluation of the most cored P25/P26 system. However, a clear systematic tension exists between the NFW baseline and both the hydrodynamical simulations and observations above $M_{\star} \sim 10^6\,\rm M_\odot$. In this regime, the profiles deviate from NFW cusps towards flatter, cored distributions. For the satellite population, this shallower signature is only uniformly preserved above $\sim 10^7\,\rm M_\odot$, as lower mass satellites present larger scatter consistent with environmental disruption.

\subsection{Direct profile comparison}
\label{sec:profile-comparison}

\begin{figure*}
        \centering
        \includegraphics[width=0.97\linewidth]{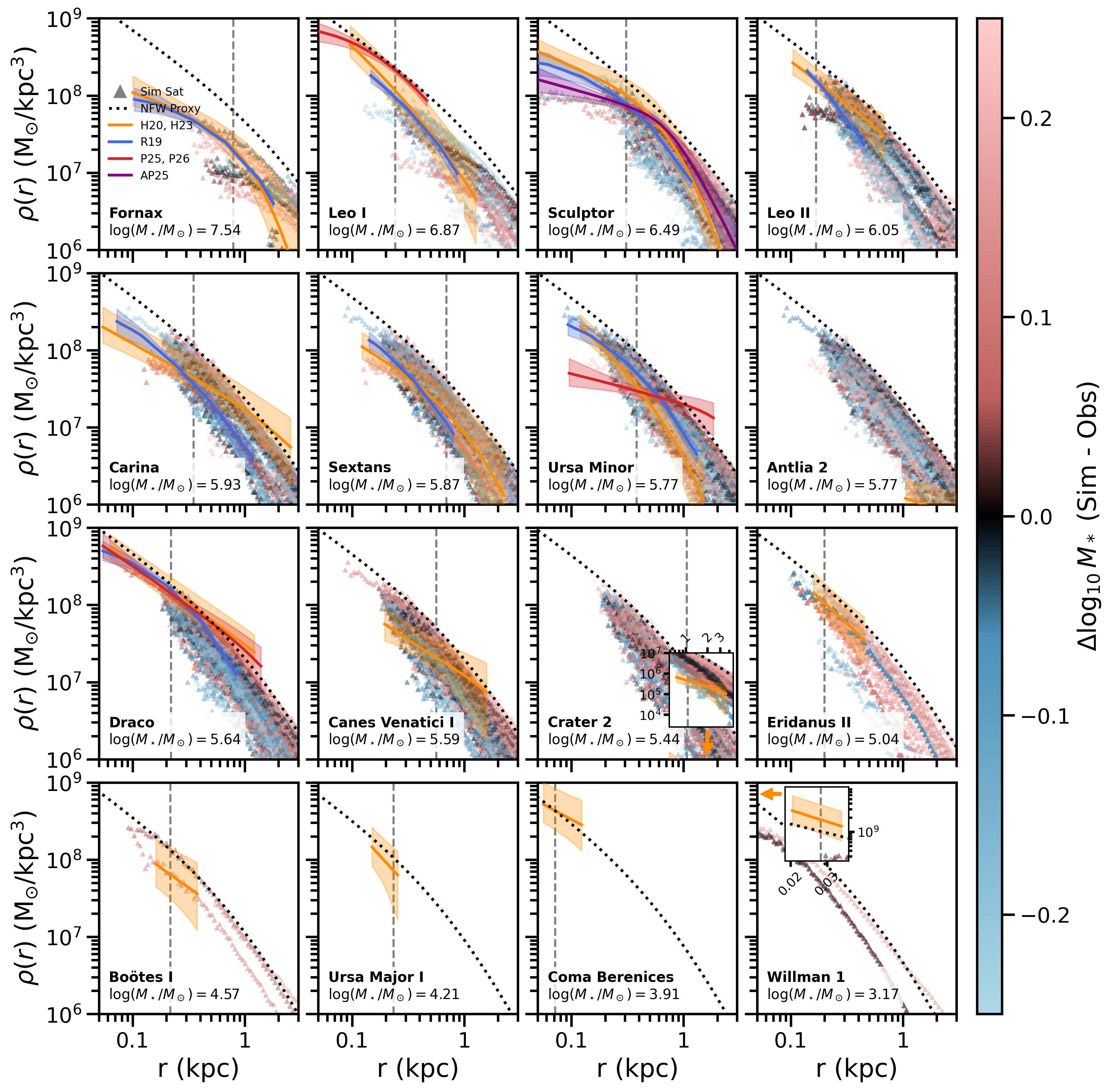}
        \caption{Dark matter density profiles versus physical radius. Each panel corresponds to one observed dwarf galaxy and includes all considered literature dynamical model estimates for that galaxy as well as the simulated galaxies whose stellar masses lie within $\pm0.25$ dex of the observed value. Dynamical model density profiles are shown as solid lines (median) with shaded bands (68\% credible intervals), colour‑coded by source: orange for H20/H23, blue for R19, purple for AP25, and red for P25/P26. Simulated satellite galaxy profiles appear as sequences of triangle symbols from their radial bins, coloured by the logarithmic deviation of the simulated stellar mass from that of the observed galaxy: black indicates the closest match, blue marks simulations with lower stellar mass, and red those with higher stellar mass. Black dotted lines represent the NFW proxy profiles derived from the \citet{Moster_2013} and \citet{Moline_2023} relations evaluated at the stellar mass of the observed galaxy. A grey vertical dashed line indicates the observed circularised projected half‑light radius. Inset panels (when present) display dynamical models whose radial extent falls largely outside the shared axis limits of the main panel. Only radial ranges that satisfy the reliability criteria for both simulations and observations are drawn.}
        \label{fig:joint-dens-phys-ms}
    \end{figure*}

\begin{figure*}
    \centering
    \includegraphics[width=0.97\linewidth]{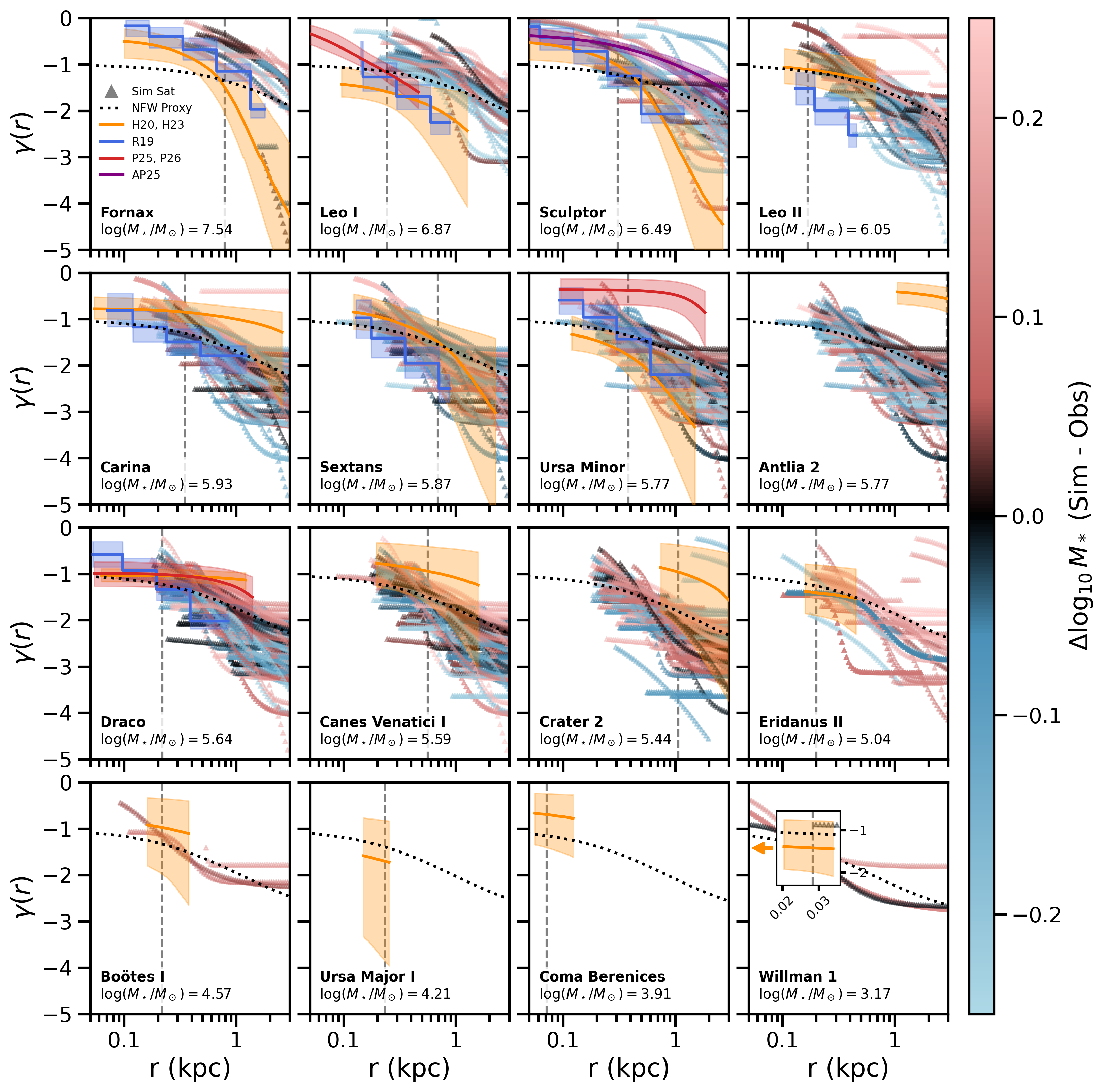}
    \caption{Same layout as Fig. \ref{fig:joint-dens-phys-ms}, but showing the logarithmic density slope versus physical radius. Satellite simulation slopes are obtained from the double power‑law fits (Eq. \ref{eq:dpl}).}
    \label{fig:joint-slope-phys-ms}
\end{figure*}

The comparison of slopes at fixed radii in the previous section summarises the profile information into a single number per galaxy per radius, which facilitates identifying trends but discards information about the entire radial structure. We complement it with a direct comparison of density and slope profiles as a function of physical radius for each of the 16 observed galaxies in the cited literature works (Fig.~\ref{fig:joint-dens-phys-ms} and \ref{fig:joint-slope-phys-ms}), restricting this primary analysis strictly to simulated satellites to ensure a physically consistent comparison that accounts for environmental effects, while keeping the comparison with centrals separate in Appendix~\ref{sec:appendix-centrals}. For the satellites, most of the simulated profiles in these figures come from FIRE-2, which provides the higher-resolution simulations in our sample.

For each observational target, we select a comparison sample of simulated satellite haloes drawn from a stellar mass window of $\pm 0.25$\,dex around the target's observed stellar mass. In both figures, simulated profiles are colour-coded by their logarithmic stellar mass offset ($\Delta\log_{10} M_{\star} = \log_{10} M_{\star,\rm sim} - \log_{10} M_{\star,\rm obs}$) using a blue-black-red colour palette, where black indicates the closest mass match. To ensure a robust comparison, we restrict all profiles to their respective reliable radial ranges, enforcing the $r_{\rm conv,20}$ convergence threshold for simulations (Sec.~\ref{sec:sim-inner-slopes}) and the data-driven spatial limits for observations (Sec.~\ref{Sec:literatureestimations}). Each panel also displays the corresponding abundance-matched NFW proxy profile as a baseline reference (Sec.~\ref{sec:NFWbaseline}), along with a vertical dashed line marking the observed half-light radius. For panels where the 
radial coverage of the observational models extends significantly beyond the shared axis limits, inset panels are included to preserve the entire unclipped profiles. Appendix~\ref{sec:appendix-profiles} (Fig.~\ref{fig:joint-dens-deviation} and \ref{fig:joint-slope-deviation}) presents the corresponding error-normalised deviations between these observationally derived profiles and their matching simulated satellite counterparts, while Appendix~\ref{sec:appendix-centrals} provides the complementary comparison with simulated centrals alone. Below, we report the main findings of this extended comparison.

\subsubsection{Density profiles: cases of good agreement}

The main result from the direct density comparison (Fig.~\ref{fig:joint-dens-phys-ms} and \ref{fig:joint-dens-deviation}) is that the observed density profiles sit inside the cloud of mass-matched simulation profiles for the large majority of galaxies, with no systematic offset in amplitude or radial shape. This is confirmed quantitatively by the deviation plots (Fig.~\ref{fig:joint-dens-deviation}): the best-matching simulations reproduce the H20/H23 profiles to within $\lesssim 1\sigma$ across essentially the full reliable radial range for most systems, and the associated 68\% envelope of best-matching profiles for each individual galaxy (see Appendix~\ref{sec:appendix-profiles}) only exceeds the $2\sigma$ deviation for Antlia\,2, Crater\,2, and for a marginal fraction of the radial span of Fornax and Sculptor. The NFW proxy (dotted lines) systematically over-predicts the central density for the more cored systems (most visibly Sculptor, Fornax, Sextans and Carina), with deviations exceeding $4\sigma$ at the smallest resolved radii, confirming that the hydrodynamical simulations provide a better description of those profiles than a pure dark-matter expectation.

Although the R19 profiles also fall broadly in the same space as simulated density profiles, their tighter reported uncertainties make them harder to match. Specifically, for several systems (most notably Fornax, Sculptor, Leo I and Leo II) the best-matching simulation deviates from the R19 median by $\gtrsim 2-3\sigma$ over a significant fraction of the radial range. We note, however, that because the R19 credible intervals are generally narrower than those of H20/H23, this pronounced $\sigma$-deviation reflects the tighter observational constraints rather than a larger absolute density offset. Furthermore, as mentioned above, the R19 profiles are the steepest ones at 250-500\,pc, and some tension may be partially attributed to dataset-to-dataset and model-to-model differences rather than a genuine physical discrepancy. We also note that the tight uncertainties of the R19 sample result in several noise spikes in Fig.~\ref{fig:joint-dens-deviation}, as they cause simulation density profile fluctuations to appear amplified.

\subsubsection{Density profiles: notable outliers}

Willman\,1 stands out as difficult to match for both simulation suites. It is the only galaxy which shows a systematically over-dense profile compared to simulations in a similar mass range, even accounting for uncertainties in the dynamical model. Although still marginally consistent with a NFW profile within $1\sigma$, the density inferred by the dynamical model available for this system is notably higher than that of a NFW cusp. Several factors complicate the comparison between simulations and observations for Willman\,1. First, statistics are limited: only four simulated galaxies fall within $\pm0.25$\,dex of its stellar mass, and none reproduce its exceptionally compact half-light radius of $\sim 28$\,pc \citep{Munoz2018}. Instead, these simulated satellites exhibit much larger half-light radii between 150 and 600\,pc. Second, the observational data are highly constrained; the H20/H23 dynamical model relies on the smallest dataset in our sample, comprising just 40 stars. Furthermore, \citet{ArroyoPolonio2026} found that the inferred dynamical mass of Willman\,1 decreases by approximately 20\% when including a statistical correction for undetected binary stars in the models, a correction missing from the H20/H23 inference. Not only does this galaxy present properties that are difficult to match by simulations, but it has also caused discussion about its classification as a galaxy. While recent studies reinforce its galaxy status \citep{Chiu2026}, historically other authors have debated whether this system is a satellite galaxy or an anomalous star cluster \citep[e.g.,][]{Willman2005, Siegel2008, Willman2011}.

At the other end, Antlia\,2 and Crater\,2 display central surface densities well below both the NFW proxy and the bulk of the mass-matched simulation sample. For Crater\,2, the downward arrow and inset panel in Fig.~\ref{fig:joint-dens-phys-ms} indicate that the central density inferred from the single available H20/H23 profile lies below $10^6\,\rm M_\odot\,kpc^{-3}$ at the smallest accessible radius, a value only marginally reached by the simulated galaxies in the matching mass window. Although Fig.~\ref{fig:joint-dens-deviation} shows the single best-matching simulations deviate only within $2\sigma$ for Antlia\,2 and $1\sigma$ for Crater\,2 across their full radial ranges, the large deviations of the 68\% best-matching simulations indicate these under-dense profiles are notably uncommon in our simulations at these stellar masses. These two objects are plausibly among the most tidally evolved satellites in the Local Group \citep{Ji2021}, and their central under-densities are tightly linked to their exceptionally large half-light radii: both systems are remarkably extended and diffuse for their stellar mass, far more than most simulated counterparts. For reference, in the literature, some simulations of Crater\,2 starting with a cored DM distribution have managed to explain the extension and dynamics of the system after strong tidal stripping by the MW \citep{Sanders2018}.

Ursa Minor also merits individual attention: the P25/P26 profile sits systematically below and flatter than both the H20/H23 and R19 constraints for the inner region of the same galaxy. This is the most pronounced internal disagreement among the different literature dynamical models we compiled for any single system in our sample. We note, however, that this discrepancy appears to depend, at least in part, on the modelling assumptions. The spherical model presented in P25/P26 is in much better agreement with the spherical inference of R19 shown here, whereas our comparison uses the axisymmetric P25/P26 model, which is more directly comparable to the flattened H20/H23 models. At the same time, \citet{Yang2025} also analysed Ursa Minor adopting flattened dynamical models and recovered a shallow density profile, more consistent with the P25/P26 axisymmetric result. Taken together, the current spread among published inferences prevents a definitive assessment of whether simulations over- or under-predict the dark matter density of Ursa Minor at different radii.

\subsubsection{Slope profiles: larger scatter, consistent trends}

The direct slope-profile comparisons (Fig.~\ref{fig:joint-slope-phys-ms} and \ref{fig:joint-slope-deviation}) confirm the core signatures evident in Fig.~\ref{fig:comparison_slopes_hlr}, but with considerably larger scatter. For the cored systems (most clearly Fornax, Sculptor, and the P25/P26 model for Ursa Minor) the observed slope profiles flatten significantly inside the half-light radius, reaching $d\log\rho/d\log r \approx -0.3$ to $-0.7$ at $r \lesssim 200$\,pc. The simulated slope profiles reproduce this inner flattening for a subset of mass-matched galaxies, while others retain a cusp, reflecting the genuine scatter in core-formation efficiency expected from variations in star formation history and stellar feedback at fixed stellar mass.

This analysis provides additional information on the mismatch between Antlia\,2 and the simulated satellite galaxies in the same mass range. Not only is Antlia\,2 notably under-dense compared to its simulated counterparts, but it is an even clearer outlier in terms of slope. The model by H20/H23 presents a shallow slope ($\approx -0.3$) at very large radii ($>1\rm \,kpc$), where all stellar mass-matched simulated satellites are still much cuspier, with slopes $\lesssim -2$. This mismatch results in a deviation profile that reaches 4-5$\sigma$ at the innermost available radial reliability range of the associated dynamical model.

The slope deviation representation (Fig.~\ref{fig:joint-slope-deviation}) also reinforces the conclusion that the R19 dataset is the most difficult to match also in slope space. The best-matching simulations reach $\gtrsim 2\sigma$ deviation from the R19 profiles for several galaxies (most notably Leo II), whereas we find simulations consistent with H20/H23 to within $\lesssim 1\sigma$ for the same systems.

A practical note on the slope-vs-density comparison is that density space provides a substantially more discriminating test for simulation-observation comparisons than slope space. The logarithmic slope is a numerical derivative of the density profile, and both observational and simulated profiles carry significant radially correlated uncertainties that are amplified on differentiation. As a result, apparent tensions in slope space can arise from small radial shifts in the position of the core-cusp transition rather than from genuine density amplitude differences, and these can dissolve when the same profiles are compared directly in density space. We therefore recommend density-profile comparisons as the primary metric for studying the agreement between dark matter density profiles of simulations and observations.

\subsubsection{Ultra-faint dwarfs}

For the four lowest-mass systems in the sample (Boötes\,I, Ursa Major\,I, Coma Berenices, Willman\,1; $\log M_{\star} < 4.6$), the reliable regions of the observational profiles cover only a narrow radial range around the half-light radius and reach very large deviations from the main panel limits for Willman\,1 (see the inset in Fig.~\ref{fig:joint-dens-phys-ms}). The best-matching simulation for Boötes\,I is consistent with the H20/H23 dynamical model to within $\lesssim 1\sigma$ over the reliable range, and the NFW proxy likewise falls within the credible intervals. While NFW profiles are also consistent with Ursa Major\,I and Coma Berenices, our sample of simulated satellites does not include any systems close to these stellar masses. However, one simulated central survives the stellar mass cut for each of them, and in both cases it matches the observations within 1$\sigma$ (see Figs.~\ref{fig:joint-dens-phys-ms-centrals}-\ref{fig:joint-slope-deviation-centrals}). The fact that both NFW proxies and simulated galaxies match these observations well is expected given that, in this mass regime, both simulations and NFW predict cuspy profiles. These systems serve as a confirmation that stellar feedback in our simulations is not extreme enough to produce cored dark matter density profiles at radial and mass scales at which observations are still compatible with cusps.

Additional caveats on the validity of dynamical model results become more relevant for low-mass dwarfs. The observed line-of-sight velocity dispersion might be artificially inflated by an unresolved binary population, which would over-predict the inferred dynamical mass if left uncorrected \citep{ArroyoPolonio2026}. Furthermore, the difficulty to obtain spectroscopic measurements for individually resolved stars in these systems implies that dynamical models must operate with limited kinematic information. To illustrate, the datasets used in H20/H23 for inferring the dark matter distribution of these galaxies comprise only between 40 and 65 stars.

\section{Conclusions}
\label{sec:conclusions}

We have compared the dark matter density profiles of Milky Way dwarf satellite galaxies, inferred from four independent dynamical modelling programmes \citep{Hayashi_slopes_1, Hayashi_slopes_2, Read_2019, chema-dynmod, Pascale_2025, Pascale_2026}, against predictions from the NIHAO \citep{wang2015} and FIRE-2 \citep{Hopkins_2018} hydrodynamical simulation suites. The central methodological contribution of this work is based on our identification of recurring inconsistent radial definitions for comparisons of logarithmic dark matter density inner slopes in the literature. Our proposed solution is to reconsider the common practice of evaluating inner slopes in simulations at a fixed fraction of the virial radius (typically 1-2\% $R_{\rm vir}$) and rather measure them at the same physical radii accessible to the observational analyses. This ensures a genuinely apples-to-apples comparison and, as we demonstrate, significantly reduces apparent tensions between cosmological hydrodynamical simulations in a CDM context and observations. Additionally, we advocate for comparing galaxies based on their stellar masses alone rather than the stellar-to-halo mass ratio, dropping the need for estimating halo masses of observed galaxies through either abundance matching methods or the dynamical models themselves, eliminating one dependency. We also note that comparing profiles of dark matter density and its logarithmic slope over the full available radial range is crucial for drawing detailed conclusions on the agreement between simulations and observations, since they provide expanded information compared to measurements at specific radial points. Our main conclusions are as follows.

\begin{itemize}

\item Physical-radius slope measurements reveal a clear stellar mass-dependent core-formation trend. Between 100 and 500\,pc, simulated isolated and satellite galaxies exhibit a rise in inner dark matter slope from super-cuspy values ($d\log\rho/d\log r \approx -1.5$) at $M_{\star} \sim 10^5\,\rm M_\odot$ to core-like values ($d\log\rho/d\log r \approx 0$ to $-0.4$) at $M_{\star} \sim 10^8\,\rm M_\odot$, with a modest re-steepening at the highest masses, as shown in Fig.~\ref{fig:comparison_slopes_hlr}. This behaviour is consistent with feedback-driven cusp-to-core transformation peaking at intermediate stellar-to-halo mass ratios \citep[e.g.,][]{DiCintio2014, dicintio_mass_profile, Tollet, Lazar, SarratoAlos2026}.

\item We identify a lower mass threshold for core formation in our set of NIHAO and FIRE-2 simulations. Up to $M_{\star} \approx 10^6\,\rm M_\odot$, simulated isolated galaxies remain marginally compatible with NFW-like cusps, indicating that stellar feedback in this regime is insufficient to consistently remove the central dark matter cusp (Fig.~\ref{fig:comparison_slopes_hlr}). This mass scale agrees broadly with theoretical predictions for the minimum feedback energy required to unbind the inner cusp \citep[e.g.,][]{Penarrubia2012, Brook2015}.

\item Simulated satellites display wider slope scatter than centrals, with a tail towards steep slopes, as indicated by the bottom row of Fig.~\ref{fig:comparison_slopes_hlr}. The satellite population develops a significant tail towards $d\log\rho/d\log r \approx -2.5$ to $-3$ absent in the isolated sample, consistent with tidal stripping of the outer halo. Since all observed targets are satellites, this wider distribution must be used as the relevant theoretical reference. This choice reduces the statistical significance of individual slope discrepancies, showing no signs of tension in the slope-mass space. This stands in contrast to previous findings from $R_{\rm vir}$-based studies. This environmental trend is consistent with exposure to tidal disruption at fixed physical distances, where a diversity in orbital histories and central halo masses acts to split the low-mass satellite population. Under this interpretation, a clear signature of slope shallowing compared to NFW cusps is only uniformly preserved above $M_{\star} \approx 10^7\,\rm M_\odot$, where deeper potential wells shield the inner profiles.

\item Simulations and observations agree well in density space (Fig.~\ref{fig:joint-dens-phys-ms} and \ref{fig:joint-dens-deviation}). When compared directly in density profile space rather than in slope space, the simulated galaxies reproduce the observed profiles to within $\lesssim 1\sigma$ for most systems and most observational datasets. For the cored systems, the simulations provide a markedly better description than an unmodified NFW profile, which over-predicts the central density by 2--3$\sigma$. We find no systematic global tension between the simulations and observations in density space.

\item Slope-space comparisons present some scatter in observational profile reconstruction methods (Fig.~\ref{fig:joint-slope-phys-ms} and \ref{fig:joint-slope-deviation}). The four modelling programmes present differences that can reach the 2$\sigma$ level in some systems (most prominently for Leo II, Draco and Ursa Minor), comparable to the apparent offset between simulations and some observational estimates. We conclude that observational modelling systematics currently represent an additional source of uncertainty that complicates their use for constraining galaxy formation and evolution models.

\item One system resists matching: Willman\,1 presents an extremely compact half-light radius that cannot be matched by CDM simulations in our dataset (Fig.~\ref{fig:joint-dens-phys-ms}). That, together with the limited range over which we deem the H20/H23 dynamical model reliable for this system, complicates finding simulated galaxies that can reproduce its high density at similar stellar masses.

\item Extremely diffuse systems, specifically Antlia\,2 and Crater\,2, challenge standard satellite evolutionary tracks within the simulation suites we analysed. Both satellites display central dark matter densities significantly lower than the bulk of the stellar mass-matched simulated population, together with exceptionally large, extended half-light radii (Fig.~\ref{fig:joint-dens-phys-ms}). Furthermore, Antlia\,2 exhibits an anomalously shallow slope ($\approx -0.3$) extending out to large physical radii ($>1$\,kpc), a feature that remains entirely unmatched by our simulated satellite sample (Fig.~\ref{fig:joint-slope-phys-ms}). These systems are likely among the most tidally evolved structures in the Local Group \citep{Sanders2018, Ji2021}, which suggests that capturing such outliers may require simulations to resolve more extreme orbital histories, or potentially explore alternative dark matter physics.

\end{itemize}

Taken together, these results demonstrate that much of the apparent discrepancy between hydrodynamical simulations and dwarf satellite galaxy observations in the cusp-core problem can be attributed to methodological inconsistencies in how inner slopes are defined and measured, compounded by significant inter-dataset and inter-model scatter in the observational constraints. Once a consistent physical-radius measurement strategy is adopted and comparisons are performed in density space, the level of agreement is high across the full stellar mass range sampled, with the notable exception of the most compact systems. Future progress will require both higher-resolution simulations that extend convergence radii to $\lesssim 50$\,pc for the faintest dwarfs, and improved observational constraints, particularly in the innermost regions of ultra-faint dwarfs, to distinguish cores from cusps below $M_{\star} \sim 10^{5}\,\rm M_\odot$ and to better understand the evolution of such systems in diverse environments.

\begin{acknowledgements}
JSA thanks the Spanish Ministry of Economy and Competitiveness (MINECO) for support through grant P/301404 from the Severo Ochoa project CEX2019-000920-S. JMA acknowledges support from the Agencia Estatal de Investigación del Ministerio de Ciencia e Innovación (AEI-MICIN) and the European Social Fund (ESF+) under grant PRE2021-100638. JMA acknowledges support from the Agencia Estatal de Investigación del Ministerio de Ciencia, Innovación y Universidades (MCIU/AEI) under grant "FOGALERA", the European Regional Development Fund (ERDF) with reference PID2023-150319NB-C21 and PID2020-118778GB-I00 and the AEI under grant number CEX2019-000920-S. Both authors thank Kohei Hayashi, Justin Read, Raffaele Pascale, and their respective collaborators for kindly sharing the results of the dynamical models of dwarf galaxies analysed in this study. They also thank Christopher Brook, Arianna Di Cintio, Giuseppina Battaglia, Jorge Moreno and Daisuke Kawata for insightful discussions. The authors wish to acknowledge the contribution of the IAC High-Performance Computing support team and hardware facilities to the results of this research. The authors also thank the NIHAO and FIRE collaborations for providing the simulations used in this work. We thank the anonymous referee for constructive comments that improved this manuscript.
\end{acknowledgements}

\bibliographystyle{aa}
\bibliography{bib.bib}

\appendix
\section{Inner slope values of observed galaxies}
\label{sec:appendix-slope-tables}

Tables \ref{tab:slope_100pc}, \ref{tab:slope_250pc} and \ref{tab:slope_500pc} list the logarithmic dark matter density slope values derived from the dynamical models of the observed satellite galaxies, evaluated at 100, 250 and 500\,pc, respectively.

\begin{table}[htbp]
\centering
\renewcommand{\arraystretch}{1.25}
\caption{Logarithmic dark matter density profile slope $\mathrm{d}\log\rho/\mathrm{d}\log r$ evaluated at 100 pc for each observed galaxy and modelling method. Values are the median of the posterior distribution with asymmetric $1\sigma$ uncertainties (16th and 84th percentiles). Studies are: R19 \citep{Read_2019}, H20/H23 \citep{Hayashi_slopes_1, Hayashi_slopes_2}, AP25 \citep{chema-dynmod}, and P25/P26 \citep{Pascale_2025, Pascale_2026}. A superscript $\dagger$ indicates that the evaluation radius falls outside the reliability region defined for that galaxy and study.}
\label{tab:slope_100pc}
\begin{adjustbox}{max width=\columnwidth}
\begin{tabular}{lcccc}
\hline\hline
Galaxy & R19 & H20/H23 & AP25 & P25/P26 \\
\hline
Antlia 2 & \nodata & $-0.37^{+0.22}_{-0.28}$$^\dagger$ & \nodata & \nodata \\
Bo\"{o}tes I & \nodata & $-0.85^{+0.54}_{-0.76}$$^\dagger$ & \nodata & \nodata \\
Canes Venatici I & \nodata & $-0.71^{+0.42}_{-0.50}$$^\dagger$ & \nodata & \nodata \\
Carina & $-0.81^{+0.36}_{-0.35}$ & $-0.79^{+0.24}_{-0.22}$ & \nodata & \nodata \\
Coma Berenices & \nodata & $-0.74^{+0.49}_{-0.78}$ & \nodata & \nodata \\
Crater 2 & \nodata & $-0.64^{+0.41}_{-0.56}$$^\dagger$ & \nodata & \nodata \\
Draco & $-0.92^{+0.25}_{-0.24}$ & $-1.03^{+0.14}_{-0.14}$ & \nodata & $-0.99^{+0.25}_{-0.28}$ \\
Eridanus II & \nodata & $-1.34^{+0.60}_{-0.45}$$^\dagger$ & \nodata & \nodata \\
Fornax & $-0.17^{+0.12}_{-0.23}$$^\dagger$ & $-0.51^{+0.27}_{-0.36}$$^\dagger$ & \nodata & \nodata \\
Leo I & $-0.88^{+0.38}_{-0.33}$$^\dagger$ & $-1.43^{+0.38}_{-0.27}$ & \nodata & $-0.67^{+0.23}_{-0.20}$ \\
Leo II & $-1.52^{+0.34}_{-0.35}$$^\dagger$ & $-1.06^{+0.40}_{-0.35}$$^\dagger$ & \nodata & \nodata \\
Sculptor & $-0.43^{+0.22}_{-0.26}$ & $-0.62^{+0.34}_{-0.33}$ & $-0.44^{+0.13}_{-0.14}$ & \nodata \\
Sextans & $-0.97^{+0.38}_{-0.44}$$^\dagger$ & $-0.82^{+0.39}_{-0.39}$$^\dagger$ & \nodata & \nodata \\
Ursa Major I & \nodata & $-1.49^{+0.77}_{-1.31}$$^\dagger$ & \nodata & \nodata \\
Ursa Minor & $-0.59^{+0.28}_{-0.30}$ & $-1.30^{+0.42}_{-0.35}$$^\dagger$ & \nodata & $-0.37^{+0.24}_{-0.31}$ \\
Willman 1 & \nodata & $-1.61^{+0.73}_{-1.57}$$^\dagger$ & \nodata & \nodata \\
\hline
\end{tabular}
\end{adjustbox}
\end{table}

\begin{table}[htbp]
\centering
\renewcommand{\arraystretch}{1.25}
\caption{Dark matter density profile logarithmic slope $\mathrm{d}\log\rho/\mathrm{d}\log r$ as Table~\ref{tab:slope_100pc}, but evaluated at 250 pc.}
\label{tab:slope_250pc}
\begin{adjustbox}{max width=\columnwidth}
\begin{tabular}{lcccc}
\hline\hline
Galaxy & R19 & H20/H23 & AP25 & P25/P26 \\
\hline
Antlia 2 & \nodata & $-0.37^{+0.22}_{-0.27}$$^\dagger$ & \nodata & \nodata \\
Bo\"{o}tes I & \nodata & $-1.00^{+0.64}_{-1.09}$ & \nodata & \nodata \\
Canes Venatici I & \nodata & $-0.80^{+0.46}_{-0.52}$ & \nodata & \nodata \\
Carina & $-1.49^{+0.28}_{-0.29}$ & $-0.82^{+0.23}_{-0.21}$ & \nodata & \nodata \\
Coma Berenices & \nodata & $-0.92^{+0.60}_{-1.35}$$^\dagger$ & \nodata & \nodata \\
Crater 2 & \nodata & $-0.71^{+0.44}_{-0.61}$$^\dagger$ & \nodata & \nodata \\
Draco & $-1.34^{+0.20}_{-0.23}$ & $-1.04^{+0.14}_{-0.13}$ & \nodata & $-1.02^{+0.25}_{-0.27}$ \\
Eridanus II & \nodata & $-1.43^{+0.61}_{-0.52}$ & \nodata & \nodata \\
Fornax & $-0.40^{+0.20}_{-0.24}$ & $-0.66^{+0.27}_{-0.29}$ & \nodata & \nodata \\
Leo I & $-1.28^{+0.31}_{-0.30}$ & $-1.60^{+0.28}_{-0.31}$ & \nodata & $-1.20^{+0.19}_{-0.19}$ \\
Leo II & $-2.00^{+0.29}_{-0.30}$ & $-1.17^{+0.40}_{-0.38}$ & \nodata & \nodata \\
Sculptor & $-1.25^{+0.20}_{-0.23}$ & $-0.91^{+0.32}_{-0.26}$ & $-0.57^{+0.10}_{-0.11}$ & \nodata \\
Sextans & $-1.41^{+0.37}_{-0.42}$ & $-1.00^{+0.35}_{-0.32}$ & \nodata & \nodata \\
Ursa Major I & \nodata & $-1.72^{+0.89}_{-2.22}$ & \nodata & \nodata \\
Ursa Minor & $-0.96^{+0.27}_{-0.27}$ & $-1.56^{+0.31}_{-0.30}$ & \nodata & $-0.37^{+0.24}_{-0.31}$ \\
Willman 1 & \nodata & $-1.82^{+0.82}_{-2.53}$$^\dagger$ & \nodata & \nodata \\
\hline
\end{tabular}
\end{adjustbox}
\end{table}

\begin{table}[htbp]
\centering
\renewcommand{\arraystretch}{1.25}
\caption{Dark matter density profile logarithmic slope $\mathrm{d}\log\rho/\mathrm{d}\log r$ as Table~\ref{tab:slope_100pc}, but evaluated at 500 pc.}
\label{tab:slope_500pc}
\begin{adjustbox}{max width=\columnwidth}
\begin{tabular}{lcccc}
\hline\hline
Galaxy & R19 & H20/H23 & AP25 & P25/P26 \\
\hline
Antlia 2 & \nodata & $-0.38^{+0.21}_{-0.27}$$^\dagger$ & \nodata & \nodata \\
Bo\"{o}tes I & \nodata & $-1.18^{+0.75}_{-1.96}$$^\dagger$ & \nodata & \nodata \\
Canes Venatici I & \nodata & $-0.92^{+0.52}_{-0.65}$ & \nodata & \nodata \\
Carina & $-1.79^{+0.22}_{-0.37}$ & $-0.89^{+0.22}_{-0.24}$ & \nodata & \nodata \\
Coma Berenices & \nodata & $-1.12^{+0.74}_{-2.28}$$^\dagger$ & \nodata & \nodata \\
Crater 2 & \nodata & $-0.79^{+0.49}_{-0.76}$$^\dagger$ & \nodata & \nodata \\
Draco & $-2.02^{+0.15}_{-0.16}$ & $-1.06^{+0.14}_{-0.14}$ & \nodata & $-1.09^{+0.26}_{-0.30}$ \\
Eridanus II & \nodata & $-1.55^{+0.64}_{-0.86}$$^\dagger$ & \nodata & \nodata \\
Fornax & $-0.68^{+0.25}_{-0.22}$ & $-1.01^{+0.30}_{-0.32}$ & \nodata & \nodata \\
Leo I & $-1.70^{+0.27}_{-0.28}$ & $-1.83^{+0.36}_{-0.71}$ & \nodata & $-1.66^{+0.24}_{-0.38}$ \\
Leo II & $-2.53^{+0.25}_{-0.25}$$^\dagger$ & $-1.30^{+0.44}_{-0.60}$ & \nodata & \nodata \\
Sculptor & $-2.07^{+0.18}_{-0.19}$ & $-1.47^{+0.29}_{-0.35}$ & $-0.75^{+0.10}_{-0.11}$ & \nodata \\
Sextans & $-1.95^{+0.35}_{-0.32}$ & $-1.30^{+0.38}_{-0.54}$ & \nodata & \nodata \\
Ursa Major I & \nodata & $-1.99^{+1.05}_{-2.81}$$^\dagger$ & \nodata & \nodata \\
Ursa Minor & $-1.43^{+0.24}_{-0.29}$ & $-1.98^{+0.41}_{-0.61}$ & \nodata & $-0.39^{+0.25}_{-0.31}$ \\
Willman 1 & \nodata & $-2.14^{+1.03}_{-3.15}$$^\dagger$ & \nodata & \nodata \\
\hline
\end{tabular}
\end{adjustbox}
\end{table}

\section{Deviations in comparisons to individual density and slope profiles of satellite galaxies}
\label{sec:appendix-profiles}

\begin{figure*}
    \centering
    \includegraphics[width=\linewidth]{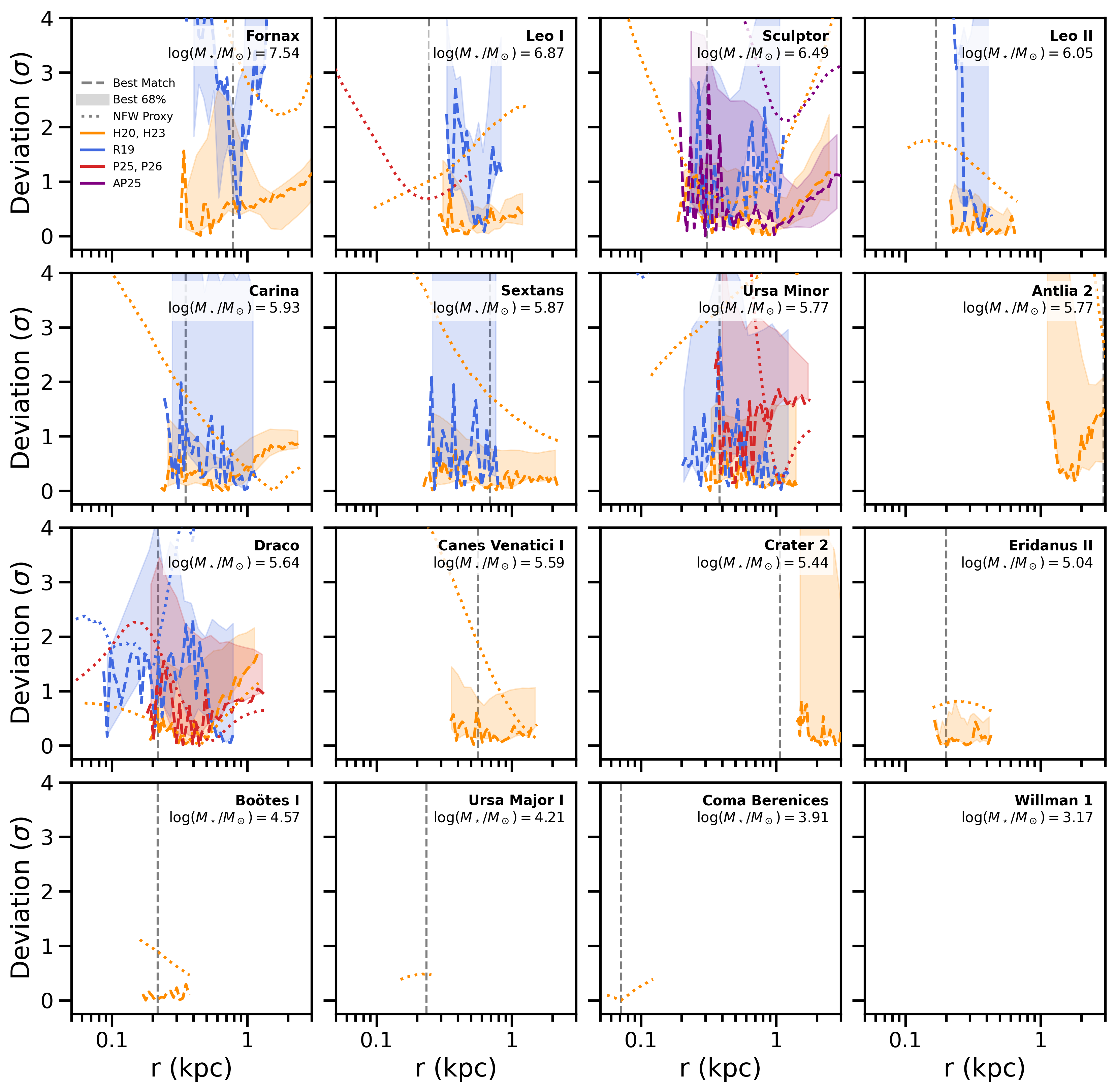}
    \caption{Deviation of simulated satellites' dark matter density profiles from individual dynamical model inferences, expressed in units of each model’s $1\sigma$ credible interval. Each panel corresponds to one observed dwarf galaxy. For each dynamical model available for that galaxy (colour‑coded by source: orange for H20/H23, blue for R19, purple for AP25, and red for P25/P26), a dashed line in the source colour shows the deviation of the mass‑matched satellite simulation that minimises the mean absolute deviation relative to that model’s median density profile; shaded bands enclose the 68\% of mass‑matched satellite simulations with the smallest mean absolute deviations from the same model; and a dotted line in the source colour gives the deviation of the NFW proxy profile derived from \citet{Moster_2013} and \citet{Moline_2023}. A grey vertical dashed line marks the observed half‑light radius. Only radial ranges deemed reliable for both simulations and observations are shown.}
    \label{fig:joint-dens-deviation}
\end{figure*}

\begin{figure*}
    \centering
    \includegraphics[width=\linewidth]{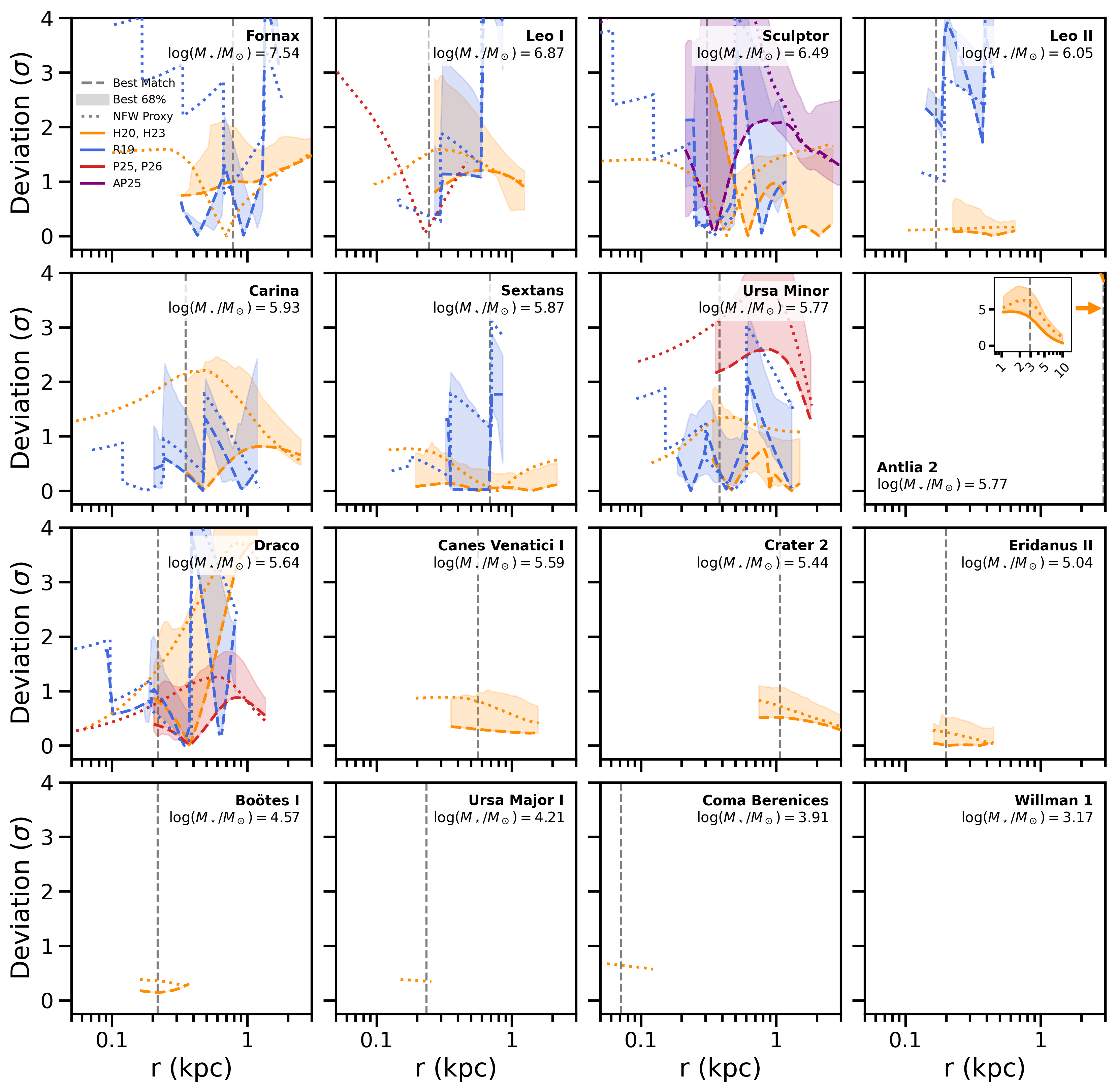}
    \caption{Same as Fig. \ref{fig:joint-dens-deviation}, but for the logarithmic density slope.}
    \label{fig:joint-slope-deviation}
\end{figure*}

This appendix complements the galaxy-by-galaxy comparisons presented in the main text (Fig.~\ref{fig:joint-dens-phys-ms} and \ref{fig:joint-slope-phys-ms}) by showing the error-normalised structural deviations between the observationally derived profiles and their matching simulated satellite counterparts (Fig.~\ref{fig:joint-dens-deviation} and \ref{fig:joint-slope-deviation}), expressed in units of each observational model's $1\sigma$ credible interval. The quantitative matching process used to pair the mass-selected simulated satellite haloes with their corresponding observational targets is detailed below. Galaxies remain ordered by decreasing stellar mass from top-left to bottom-right.

The profile comparisons in Fig.~\ref{fig:joint-dens-deviation} and \ref{fig:joint-slope-deviation} are performed separately for each observational source available for a given galaxy. For each source, we identify the best-matching simulation as the mass-matched satellite galaxy that minimises the mean error-normalised absolute deviation ($\text{MAD}_{\sigma}$) over the overlapping reliable radial range, which is defined as:

\begin{equation}
\label{eq:mad_profiles}
\text{MAD}_{\sigma} = \frac{1}{N_{\rm pts}} \sum_{i=1}^{N_{\rm pts}} \delta_i\,,
\end{equation}

\noindent where $N_{\rm pts}$ is the number of overlapping radial points, and the dimensionless residual $\delta_i$ at each radial coordinate $r_i$ is defined by scaling the absolute difference by the asymmetric $1\sigma$ observational uncertainty envelope:

\begin{equation}
\delta_i = \begin{cases} 
\dfrac{y_{\rm sim}(r_i) - y_{\rm obs, med}(r_i)}{y_{\rm obs, h68}(r_i) - y_{\rm obs, med}(r_i)} & \text{if } y_{\rm sim}(r_i) > y_{\rm obs, med}(r_i)\,, \\ 
\dfrac{y_{\rm obs, med}(r_i) - y_{\rm sim}(r_i)}{y_{\rm obs, med}(r_i) - y_{\rm obs, l68}(r_i)} & \text{if } y_{\rm sim}(r_i) \le y_{\rm obs, med}(r_i)\,. 
\end{cases}
\end{equation}

\noindent Here, $y_{\rm sim}$ represents the simulated profile value (either density $\rho$ or logarithmic slope $d\log\rho/d\log r$), while $y_{\rm obs, med}$, $y_{\rm obs, l68}$, and $y_{\rm obs, h68}$ represent the median, $16^{\rm th}$, and $84^{\rm th}$ percentiles of the observational credible interval, respectively. Simulated satellite galaxies are only considered valid candidates if the overlapping spatial segment covers at least 50\% of the total logarithmic width of the observational reliable range:

\begin{equation}
\frac{\log_{10}(r_{\rm ov, max} / r_{\rm ov, min})}{\log_{10}(r_{\rm obs, max} / r_{\rm obs, min})} \ge 0.5\,,
\end{equation}

\noindent where $r_{\rm ov, min}$ and $r_{\rm ov, max}$ are the minimum and maximum radii of the radial segment over which the observational and simulated reliable ranges overlap [i.e. $r_{\rm ov, min} = \max(r_{\rm obs, min}, r_{\rm conv,20})$ and $r_{\rm ov, max} = \min(r_{\rm obs, max}, R_{\rm vir})$].

Shaded bands enclose the 68\% of mass-matched satellite simulations with the lowest $\text{MAD}_{\sigma}$ values relative to the observed median, providing a measure of the structural spread among the best-fitting simulated profiles. For reference, the dotted line in the source colour shows the deviation of the NFW proxy profile from the observed median. A deviation of zero indicates perfect agreement with the observed median; a value of $1$ corresponds to the $1\sigma$ boundary of the observational credible interval.

\section{Comparison with simulated central galaxies alone}
\label{sec:appendix-centrals}

For completeness, this appendix presents the comparison of the 16 observed dwarf satellites exclusively against simulated central galaxies.  

The simulated central profiles are selected using an identical stellar mass window of $\pm 0.25$\,dex around each target's observed stellar mass. The optimisation process to define the best-matching haloes and the 68\% minimum-deviation envelopes follows the exact mathematical formulation described in Appendix~\ref{sec:appendix-profiles}. 

\begin{figure*}
    \centering
    \includegraphics[width=\linewidth]{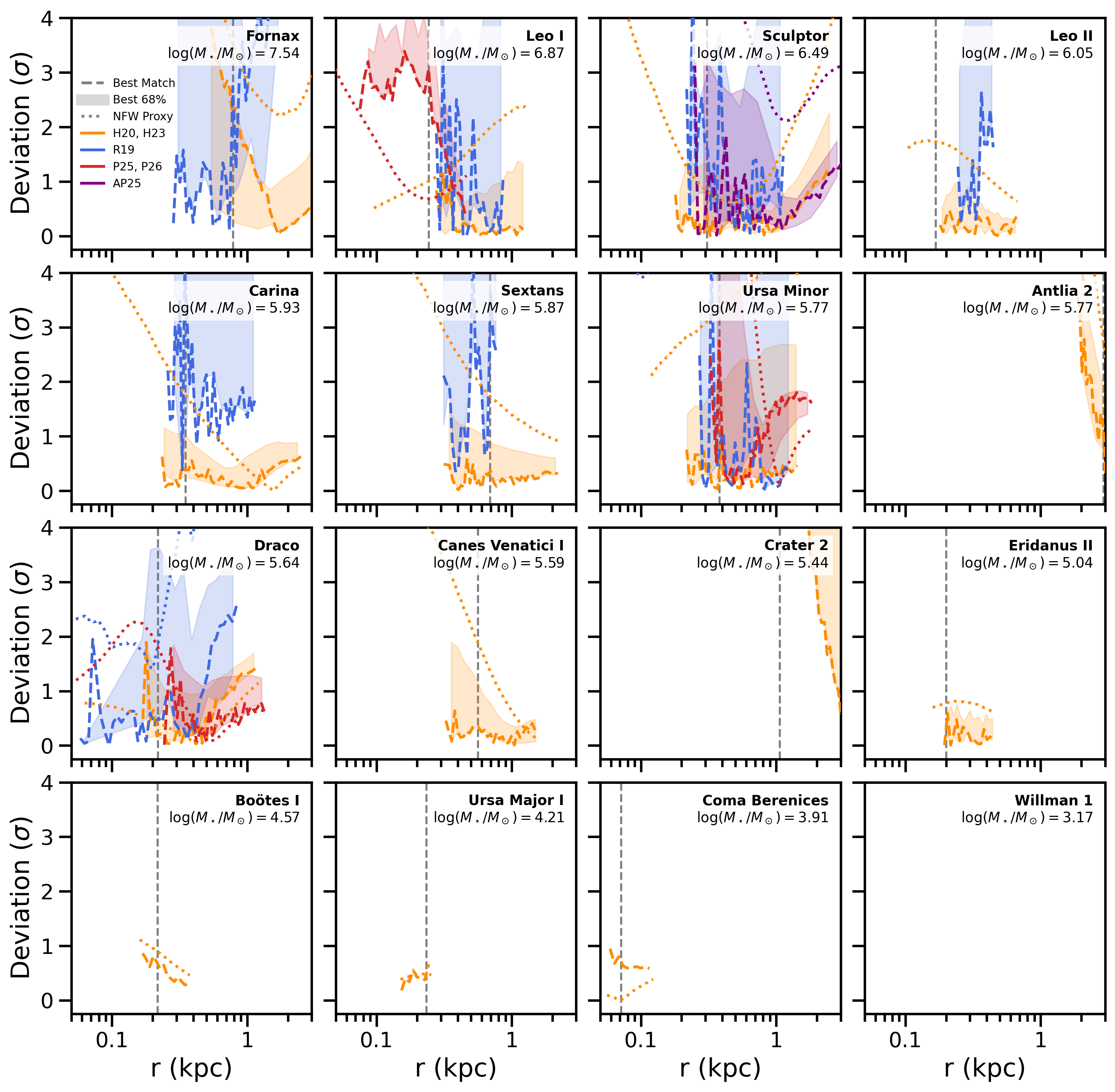}
    \caption{Deviation of simulated central dark matter density profiles from individual dynamical model inferences, expressed in units of each model's $1\sigma$ credible interval. Layout, symbols, and literature colour-coding are identical to Fig.~\ref{fig:joint-dens-deviation}, but simulated central profiles are represented here as circular points.}
    \label{fig:joint-dens-deviation-centrals}
\end{figure*}

\begin{figure*}
    \centering
    \includegraphics[width=\linewidth]{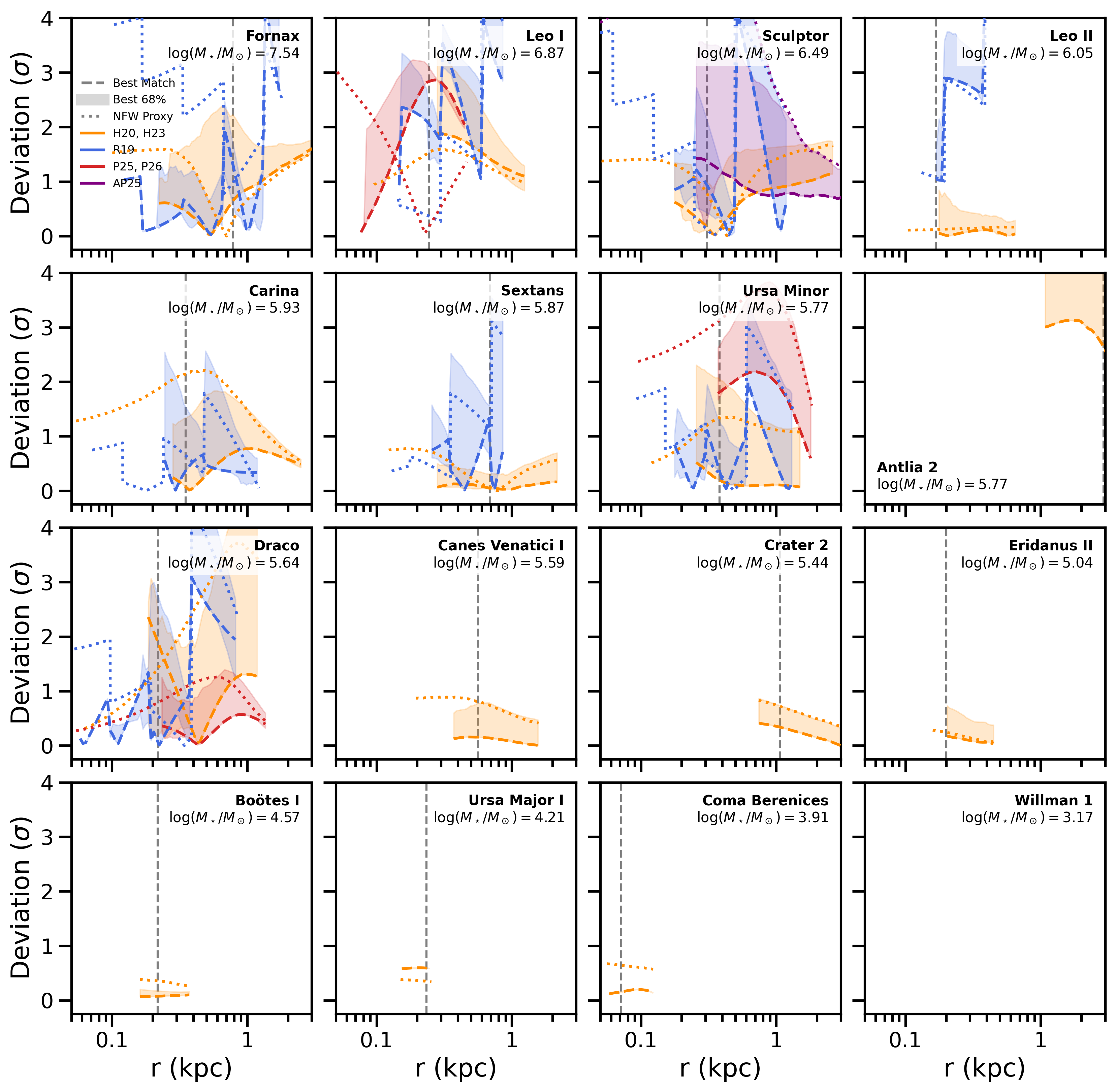}
    \caption{Same as Fig.~\ref{fig:joint-dens-deviation-centrals}, but for the logarithmic density slope.}
    \label{fig:joint-slope-deviation-centrals}
\end{figure*}

Fig.~\ref{fig:joint-dens-phys-ms-centrals} presents the comparison for density profiles and Fig.~\ref{fig:joint-slope-phys-ms-centrals} for slope profiles. Figs.~\ref{fig:joint-dens-deviation-centrals} and \ref{fig:joint-slope-deviation-centrals} display their respective error-normalised deviation profiles.

While these comparisons are hindered by noise spikes amplified due to error normalisation of deviations, we highlight that density profile deviations with respect to centrals are in general larger than with respect to simulated satellites. As an example, the R19 deviations with respect to best-matching profiles exceed 3$\sigma$ for several cases (Leo I, Sculptor, Carina, Sextans, Ursa Minor), while in the comparison to satellites (Fig.~\ref{fig:joint-dens-deviation}) they remain below that value (and also below 2$\sigma$ in the cases of Carina and Sextans) over the full reliability region of the dynamical model.

\begin{figure*}
    \centering
    \includegraphics[width=\linewidth]{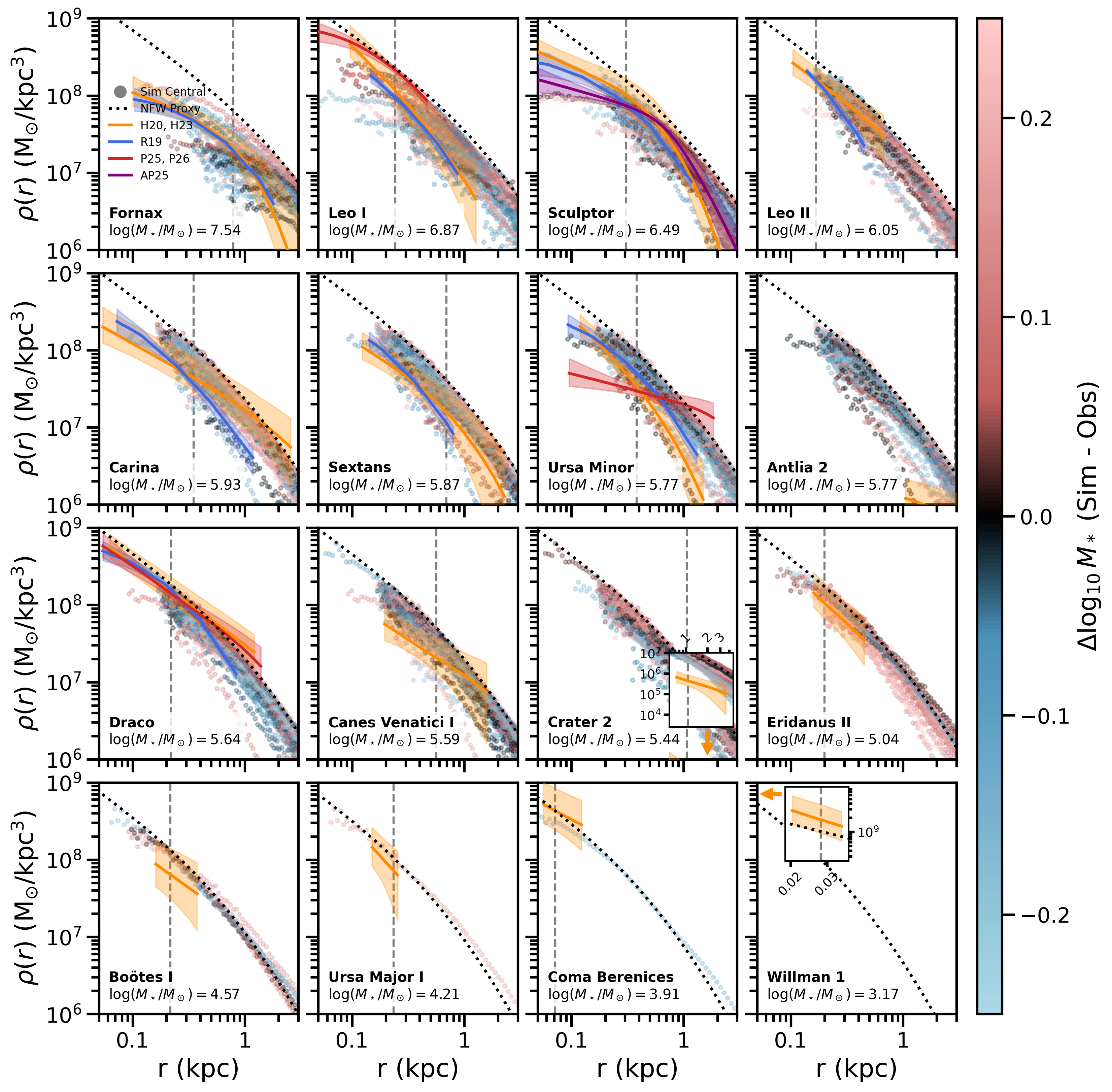}
    \caption{Dark matter density profiles versus physical radius, comparing observed satellite galaxies exclusively with simulated central galaxies. Layout, symbols, and literature colour-coding are identical to Fig.~\ref{fig:joint-dens-phys-ms}, but simulated central profiles are represented here as circular points.}
    \label{fig:joint-dens-phys-ms-centrals}
\end{figure*}

\begin{figure*}
    \centering
    \includegraphics[width=\linewidth]{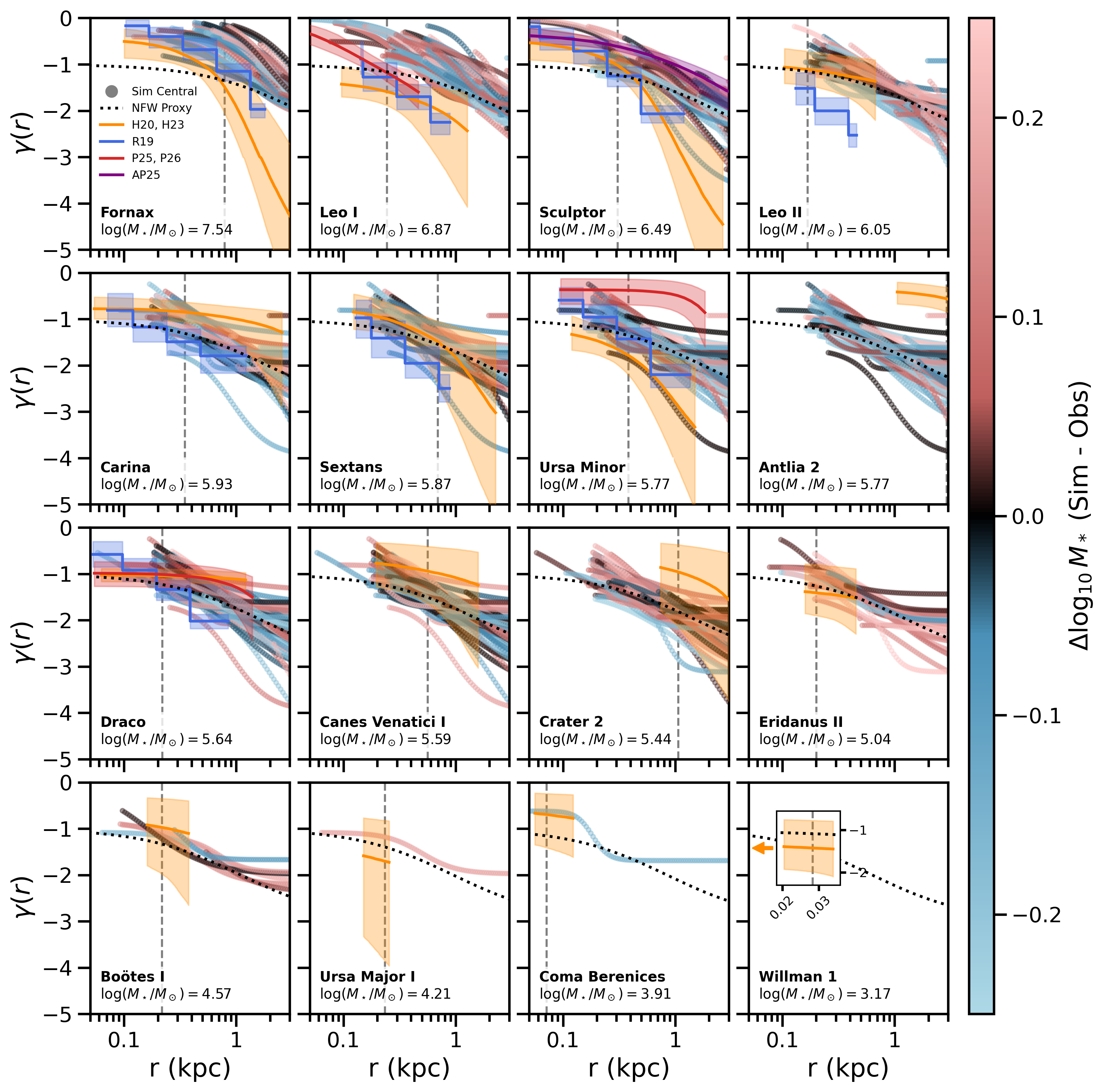}
    \caption{Same as Fig.~\ref{fig:joint-dens-phys-ms-centrals}, but for the logarithmic density slope.}
    \label{fig:joint-slope-phys-ms-centrals}
\end{figure*}

\label{LastPage}
\end{document}